\documentclass[prd,eqsecnum,floats,nopacs]{revtex4}
\usepackage{amsmath}
\usepackage{amssymb}
\usepackage{amsfonts}
\usepackage{dsfont}

\newcommand{\bs}[2]{{{\sf{b}}
\phantom{]}}^{\!\!\!\!\!\!*{\kern 1.3pt}{\mbox{${\scriptstyle #1}$}}}_{\mbox{${\scriptstyle #2}$}}}
\newcommand{\as}[2]{{{\sf{a}}
\phantom{]}}^{\!\!\!\!\!\!*{\kern 1.3pt}{\mbox{${\scriptstyle #1}$}}}_{\mbox{${\scriptstyle #2}$}}}
\newcommand{\naf}[2]{n'^{\!\!\!\!\!*{\kern 2.8pt}{\mbox{${\scriptstyle(#1)}$}}}
_{\mbox{${\scriptstyle #2}$}}}
\newcommand{\na}[2]{n^{\!\!\!\!*{\kern 1.3pt}{\mbox{${\scriptstyle(#1)}$}}}
_{\mbox{${\scriptstyle #2}$}}}
\renewcommand{\ap}[2]{a^{\!\!\!\!*{\kern 1.3pt}{\mbox{${\scriptstyle(#1)}$}}}
_{\mbox{${\scriptstyle #2}$}}}
\newcommand{\ep}[1]{e^{\!\!\!\!*{\kern 1.3pt}{\mbox{${\phantom{()}}$}}}
_{\mbox{${\scriptstyle #1}$}}}

\newcommand{\ad}[2]{a^{\!\!\!\!*{\kern 1.3pt}{\mbox{${\scriptstyle #1}$}}}_{\mbox{${\scriptstyle #2}$}}}

\newcommand{\gug}[3]{{#1}^{\!\!\!\!*{\kern 1.3pt}{\mbox{${\scriptstyle #2}$}}}_{\mbox{${\scriptstyle #3}$}}}

\newcommand {\oks}[2]{{\raise0.7ex\hbox{${\scriptstyle #1}$}\!\mathord{\left/
{\vphantom{{1}{2}}}\right.\kern-\nulldelimiterspace}
\!\lower0.7ex
\hbox{${\scriptstyle #2}$}}}

\newcommand{\ii}{\mathrm{i}}

\begin{document}

\title{Particle quantum states with indefinite mass and neutrino oscillations}

\author{A. E. Lobanov}\email{lobanov@phys.msu.ru}

\affiliation {Department of Theoretical Physics, Faculty of Physics,
  Moscow State University, 119991 Moscow, Russia}

\begin{abstract}
In this work we develop a mathematical formalism which allows obtaining oscillation formula for neutrino of any energy. We demonstrate that in the ultra-relativistic limit the results obtained in this new approach agree with the previously used phenomenological theory which is only applicable to ultra-relativistic neutrinos. To this end we do the following.
The Hilbert spaces of particle states are constructed
in such a way that the neutrinos are combined in a  multiplet with
its components being considered as different quantum states
of a single particle. The same is done for the  charged leptons
and the down- and up-type quarks. In  the  theory based on
the Lagrangian of the fermion sector of the Standard Model
modified in accordance with this construction, the phenomenon of
neutrino oscillations arises as a direct consequence of  the
general principles of quantum field theory. Using the example of the pion decay, when the resulting neutrino has to be ultra-relativistic, it is shown that the neutrino
states  produced in the decay process can be described by a
superposition of states with different masses and identical
canonical momenta with very high accuracy.
\end{abstract}

\maketitle

\section{Introduction}\label{0}

The Standard Model of electroweak interactions  based on the
non-Abelian gauge symmetry of  the interactions \cite{Glashow,Weinberg,Salam},
the generation of particle masses due to the  spontaneous symmetry
breaking mechanism \cite{Englert,Higgs,Kibble},  and the  philosophy of mixing of particle
generations \cite{Cabibbo,Kobayashi} is universally recognized. Its
predictions obtained in the framework of perturbation theory are
in a very good agreement with the experimental data.  There
is no serious  reason, at least at  the energies  available at
present, for its main propositions to be revised.

However, in describing such an important and  firmly
ex\-pe\-ri\-mentally established pheno\-me\-non as neutrino
oscillations (see \cite {pdg}), an essentially  phenomenological
theory   based on the pioneer works  by B. Pontecorvo
\cite{Pontecorvo} and  Z. Maki et al. \cite{MNS} is used.  Its
statements are given in various review articles and monographs
(see, e.g., \cite{bp,kayser,osc3,dolg,MP,FY,Giunti,bil,AS}). The
primary assumption of this theory is that  the neutrinos are
massive and  there are three  neutrino types with different
masses. It is also postulated that the neutrinos produced in
reactions are in the states which are superpositions of states
with fixed masses, the so-called mass  states.
These superpositions  build the so-called flavor basis. The
transformation to this basis from the mass basis is given by a
unitary mixing matrix. Initially, the mass basis elements are
described by plane waves with the same (three-dimensional)
momentum. The time evolution of the flavor states is described by
the solution of the corresponding Cauchy problem.

The conclusions of this theory  are valid when the neutrino
energy is large compared to their masses $ m_{\nu}^{2}/{\cal
E}_{\nu}^{2}\ll 1$.  The neutrino masses are extremely
small. Due to this fact,  the neutrino energies, which satisfy
this condition, are not too high. Therefore, there is an obvious
logical contradiction. On the one hand, neutrino oscillations are
observed at the energies that are typical for the Standard
Model. On the other hand, oscillations suggest the possibility of
transformations  of free fermions with equal electroweak quantum
numbers into each other,  contrary to the Standard
Model.

In the Standard Model of electroweak interactions, mass generation is due to spontaneous symmetry breaking. Mass matrices are diagonalized after spontaneous symmetry breaking. This transformation leads to the appearance of a mixing matrix in the terms of the interaction Lagrangian associated with charged currents. The mixing in the Lagrangian of free fields is absent and the fields of fermions with different masses are quantized independently. As the consequence, all fermions in this model  are  fundamental particles uncorrelated with each other.

Numerous efforts based on  quantum field theory  have
been made to justify the conclusions of the phenomenological
model. It is worth mentioning the description of the  neutrino
states in the form of wave packets (see, e.g., \cite{Giunti_x,Giunti_y}).
Another approach involves a description of the  neutrino
production and  detection as a single process (see, e.g.,
\cite{Grimus,Kiers,Grimus1,Naumov}).  A review of the models
based on the quantum theory of fields is given in paper
\cite{MB}. The results obtained
partially explain the oscillation mechanism,
however,  these models  are
not complete in the mathematical sense.

All the contradictions of the theory could be solved, if we
could combine the neutrinos in a multiplet with their components
being considered as different quantum states of a single particle.
This could be done for the  charged leptons and the down- and
up-type quarks as well. To this purpose it is necessary to
construct  the Fock spaces  of  particle  flavor states.

Attempts to construct the Fock space for neutrino flavor states by a mathematically-consistent way have been repeated many times \cite{Giunti_1,Fujii,Fujii1,Blasone,Blasone1,Blasone2,Giunti_2}. However, either the commutation relations for the creation and annihilation were differed from the canonical type \cite{Giunti_1,Fujii,Fujii1}, or the vacuum of the theory was explicitly dependent on time \cite{Blasone,Blasone1,Blasone2}. Thus it was concluded that it is likely impossible to construct the Fock space for the flavor states using the conventional approach \cite{Giunti_2}, i.e. it is not possible to construct a unitary transformation between the mass and the flavor states using  the mixing matrix only \cite{han}. For a brief historical summary of the problem see, e. g., \cite{Ho}. Note that a possibility of constructing the unitary transformation between the mass and the flavor states using the mixing matrix only was taken for granted in the phenomenological theory.

Let us consider this issue   in detail.
In relativistic quantum field theory a particle is usually
associated with an irreducible representation of the Poincar\'{e}
group \cite{W39,boloto}.  The eigenvalue  of the Casimir operator
constructed from the translation operators squared is
identified with the observed mass of the particle.
Therefore the existence of states that are super\-posi\-tions of one-particle
states with different masses contradicts the relativistic
invariance of the theory, if  the canonical momentum operator is
identified with the translation operator. This obviously follows
from the fact that the metric is defined on the hyperboloid in the
momentum space determined by the value of  the particle mass.

To overcome this difficulty it seems natural to associate  some
set of particles (a multiplet) with  an irreducible representation
of a wider group. A number of theorems
\cite{Michel,Coleman,Weinberg2}  indicate that  the only
reasonable extension of the symmetry group of the theory is the
direct product of the Poincar\'{e} group and a group of internal
symmetry. This immediately limits the set of the models under
consideration. However, if we assume that the  masses of particles
in a multiplet can be set by hands, as it is done in the Standard
Model, this circumstance is not essentially important. It is
essential that in this point we encounter the Jost  theorem
\cite{Jost,Raif,Raif1},{\footnote{Let $\Gamma$ be a continuous unitary
representation of a finite-dimensional connected Lie group $G$  in
a Hilbert space ${\cal H}$. Let $G$ contain the inhomogeneous
Lorentz group as an analytic subgroup. Let finally the spectrum of
the 4-momentum operator $P_{\mu}$ be contained in $\{0\}\bigcup\,
V^{+}, V^{+}$ being the future cone in the Minkowski space
$M^{4}$. If $m_{1}>0$ is an isolated eigenvalue of the mass
operator $M = (P_{\mu}P^{\mu})^{1/2}$ then the corresponding
eigenspace ${\cal H}_{1}$ is invariant under
$\Gamma(G)$.}}. It seems that due to this
theorem the result of extending the symmetry group of the theory
will be trivial. Since the  irreducible unitary representations of
the direct product of groups are unitary equivalent to the tensor
product of the representations of these groups with the same
values of the invariants{\footnote{As a matter of fact, the given
statement  is true for finite-dimensional representations.
However, in what follows we shall consider only the  constructions
which are unitary equivalent to  each other.}}, the masses
of the  components of the multiplets will be equal. It is
this circumstance that is the most important reason of the
failure of constructing the Fock space for flavor states.

In our opinion, it is possible to circumvent this obstacle \cite{Lob}. The
basic idea of the proposed approach is as follows. The conclusion
mentioned above is not correct if there is  more than one
multiplet in the theory, and these multiplets interact with each
other. As a matter of fact, in the general case the representation
spaces for two multiplets cannot be simultaneously reduced to the
direct sum of  representation spaces of the Poincar\'{e} group.
Even if the representation of the internal symmetry group is
considered  to be finite, which seems quite natural for a
finite set of particle masses in the multiplet,  the
representations  for both multiplets would be unitary equivalent,
but not identical. The values of the invariant  associated
with the translation operators squared  will be the same for all
 the components of the multiplets (for different multiplets
the values may vary). However, there is no need to treat them as
the observed squared masses of the particles. The masses will be
determined by an operator from the enveloping algebra for the
direct sum  of the Lie algebras of the Poincar\'{e} group and the
internal symmetry group.
Therefore, the representations
which  were described above, being unitary equivalent are not
``physically equivalent''. Such consideration is conformed to the
spirit of the Standard Model, where the masses of all  particles
are generated due to the phenomenon of spontaneous symmetry
breaking  and are proportional to the vacuum expectation value  of
the Higgs field.

The purpose of this work is to implement these ideas to construct the Fock space for the flavor states and to develop
a modification of the Standard Model, which is suitable to describe the
particle oscillations.
In such a model it will be possible to describe the behavior of not only ultra-relativistic neutrinos, but also  low-energy neutrinos, for example,  relic ones.
To do this, in Section \ref{M.1}, following the results of the article \cite{Lob}, we construct  wave functions of
multiplets of fermions with spin $1/2$, which form the spaces of
irreducible representations of the direct product of  the
Poincar\'{e} group and  the group $SU(3)$. First, we consider the
most simple version of such a construction that is the tensor
product of the Dirac  representation of the Poincar\'{e} group and
the fundamental representation of $SU(3)$, which we designate as
${\cal{H}}_{m,1/2}$. In this space the wave functions describe
multiplets with components possessing equal masses $m$.

Then,  using the example of neutrino multiplet, we find
a unitary intertwining operator mapping ${\cal{H}}_{m,1/2}$ to a
new space ${\cal{H}}_{m,1/2}^{(\nu)}$. In this space the wave
functions describe  a multiplet with components possessing
different masses that are the eigenvalues of the  operator, which
is defined by a mass matrix of the multiplet. These components
describe the so called mass states of the field. The
mass states generate a basis in
${\cal{H}}_{m,1/2}^{(\nu)}$, but the choice of this basis is not
unique. We find all the plane-wave bases, which can describe the
mass states. For any choice of the mass basis, linear
combinations of its elements are bases in
${\cal{H}}_{m,1/2}^{(\nu)}$. Due to  the unitary equivalence
of representations in ${\cal{H}}_{m,1/2}$  and in
${\cal{H}}_{m,1/2}^{(\nu)}$, the eigenvalues of the Casimir
operator constructed from the translation operators squared on the
elements of ${\cal{H}}_{m,1/2}^{(\nu)}$ are equal to  $m^{2}$.
However, the parameter $m$ is not the observed mass, it  sets the
scale of the multiplet masses. As a matter of fact, the parameter
$m$ has the meaning of  the vacuum expectation value of the
Higgs field.

In Section \ref{M.2}, using the results of Section \ref{M.1}, we
introduce  a Lagrangian generalizing the fermion sector of the
Standard Model in such a way that field functions of the
individual particles are replaced by field functions of the
multiplets. In the model under consideration the action defined by
the Lagrangian of free fields is invariant with respect to the
direct product of the Poincar\'{e} group and $SU(3)$. Using
Noether's theorem we find the integrals of motion for the fields
and carry out the quantization procedure. We make sure that
the creation and annihilation operators satisfy the canonical
commutation relations. However, these operators possess an
additional discrete quantum number that is associated with the
mass of the state. The results of the quantization make it
possible to treat both the mass states of the multiplet and their
superpositions as the quantum states  of a single particle and to
construct  the appropriate Fock space. Therefore, the transition
probabilities can be calculated in the framework of perturbation
theory. In the theory based on the Lagrangian of the fermion
sector of the Standard Model modified in accordance with this
construction, the phenomenon of particle oscillations arises.

Note again that in the Standard Model  non-diagonal matrices of
Yukawa couplings are introduced to describe the interaction of
fermions with the Higgs field. This procedure violates the
$SU(3)$-symmetry of the theory. In the framework of the
Kobayashi--Maskawa   formal\-ism \cite{Kobayashi},  the
diagonalization of these matrices is necessary to generate masses
of fermions. This transformation leads to a mixing matrix in the
charge-current interaction Lagrangian. This procedure presupposes
that all the fermions described by the model (i.e. the
neutrinos, the charged leptons and the down- and up-type quarks)
are fundamental particles. Therefore, although the mixing of
generations occurs, the direct transitions between free particles
with the same electroweak quantum numbers, i.e. oscillations,  are
impossible.

In the proposed model  the mixing occurs at the level of
perturbation theory. So there is no need  to carry out the
diagonalization procedure and to introduce the mixing matrix into
the interaction Lagrangian explicitly. It is obvious that the
transition probabilities, obtained in this way, will oscillate.
However, at least the  transition probabilities obtained in
the tree approximation will  completely coincide with the
predictions of the conventional approach at  the distances
from the source that are small compared with the  oscillation
lengths.

In the following sections we apply the results to the description
of neutrino oscillations.  In Section \ref{M.3} we check that the
formulas describing the neutrino oscillations are independent of
the type of the mass states, i.e. of the  choice of the plane-wave
basis in ${\cal{H}}_{m,1/2}^{(\nu)}$. In Section \ref{M.4} we
calculate the probability of the pion decay and check that the
probability of the production of one or another flavor state
essentially depends on the types of the mass states of which  the
flavor states are constructed. A single flavor state is produced
only when it is composed of the mass states with the same
canonical momentum. Thus, we come to the conclusion that the
phenomenological theory of high energy neutrino oscillations based
on the ideas of Pontecorvo is a very good approximation  for
quantum field theory.

\section{Spaces of wave functions}\label{M.1}

Let us assume that  the  neutrinos, the  charged leptons, the  down- and up-type quarks are elements of
different multiplets. For each multiplet
we consider a space $ {\cal{H}}_{m ,1/2} $, which is the direct
sum of the Dirac representation spaces for the Poincar\'{e} group
with fixed (positive)
frequency and equal values of the Casimir operator formed from the
canonical momentum operators squared. Since  the  experimental data indicate that there are three generations of particles,
\begin{equation}\label{01}
  {\cal{H}}_{m,1/2}={\cal{H}}_{m,1/2}^{(1)}
  \oplus{\cal{H}}_{m,1/2}^{(2)}
  \oplus{\cal{H}}_{m,1/2}^{(3)},
\end{equation}
\noindent where the  subspaces ${\cal{H}}_{m,1/2}^{(l)}$
correspond to the  components of the  multiplet.

In the  coordinate representation a basis of $
{\cal{H}}_{m, 1/2} $ can be defined as
\begin{equation}\label{001}
  \Psi_{p,\zeta,l}(x)= \psi_{p,\zeta}(x)\,e^{(l)}.
\end{equation}
\noindent Here $\psi_{p,\zeta}(x)$ are plane waves
\begin{equation}\label{0001}
  \psi_{p,\zeta}(x)=\frac{1}{\sqrt{2p^{0}}}
  u_{p,\zeta}e^{-\ii (px)},
\end{equation}
\noindent where $p^{0}=\sqrt{{\bf p}^{2}+m^{2}}$, and the
spinors $u_{p,\zeta}$ satisfy the equation
\begin{equation}\label{00010}
(\gamma^{\mu}p_{\mu}-m)u_{p,\zeta}=0.
\end{equation}
\noindent This spinors  are normalized by the condition
\begin{equation}\label{000100}
{\bar{u}}_{p,\zeta}{u}_{p,\zeta'}=
2m\delta_{\zeta,\zeta'},\quad {\bar{u}}_{p,\zeta}= {u}^{\dag}_{p,\zeta}\gamma^{0}.
\end{equation}
\noindent The  indices $ \zeta =\pm 1 $ determine  the
particle polarization.

We can consider  the space of the
representation with negative frequency, too.  A basis in this space can  be introduced as
\begin{equation}\label{001xxxx}
\begin{array}{c}
 \displaystyle \Psi_{-p,\zeta,l}(x)= \psi_{-p,\zeta}(x)\,e^{(l)},\;\; \psi_{-p,\zeta}(x)=\frac{1}{\sqrt{2p^{0}}}
 u_{-p,\zeta}e^{\ii (px)},\\ \displaystyle (\gamma^{\mu}p_{\mu}+m)u_{-p,\zeta}=0,\;\;  {\bar{u}}_{-p,\zeta}{u}_{-p,\zeta'}=
 -2m\delta_{\zeta,\zeta'}.
\end{array}
\end{equation}
\noindent The explicit formulas for the spinors $u_{p, \zeta},
{\bar{u}}_{p,\zeta} $,  $u _ {-p, \zeta}, {\bar{u}}_{-p,\zeta} $,
see, e.g., in \cite{LL}. In what follows we will work in the
spaces of wave functions with the positive frequency. The
conversions for the spaces of wave functions with the negative
frequency are similar.

Consider the vectors $e^{(l)} $ which, for definiteness, can be
chosen in the form
 \begin{equation}\label{00001}
e^{(1)}=\left(
          \begin{array}{c}
            1 \\
            0 \\
            0 \\
          \end{array}
        \right),\quad
e^{(2)}=\left(
          \begin{array}{c}
            0 \\
            1 \\
            0 \\
          \end{array}
        \right),\quad
e^{(3)}=\left(
          \begin{array}{c}
            0 \\
            0 \\
            1 \\
          \end{array}
        \right) ,
\end{equation}
\noindent as a basis of a three-dimensional vector space over
the field of complex numbers. Introduce  a scalar product
in $ {\cal{H}}_{m,1/2}$
\begin{equation}\label{000001}
({\Psi}, \Phi)= \sum_{s=1}^{3}\int\! d{{\bf
x}}\,{\Psi}^{\dag}_{s}({\bf x},t)\,\Phi_{s}({\bf x},t),
\end{equation}
\noindent with summation over the coordinates of the vectors
$e^{(l)}_{s}$. Then one can consider ${\cal{H}}_{m,1/2}$
as the  space of an irreducible unitary representation of the
direct product of the Poincar\'{e} and $SU(3)$ groups. This space is
constructed as the direct product of the Dirac representation
of  the Poincar\'{e} group and the fundamental representation of
the group $SU(3)$.

The explicit form of the Lie algebra elements in the pre\-sent case
is obvious.  We have the standard realization for the generators
of the Poincar\'{e} group
\begin{equation}\label{00000001}
 P_{\mu} = \ii{\partial_{\,\mu}}{\mathds{I}},\quad M_{\mu\nu}
=\ii\Big((x_{\mu}\partial_{\,\nu}-x_{\nu}
\partial_{\,\mu})+
(\gamma_{\mu}\gamma_{\nu}-\gamma_{\nu}
\gamma_{\mu})/4\Big)
{\mathds{I}},
\end{equation}
\noindent where  ${\mathds{I}}$ is the $3 \times 3$ identity
matrix. The Hermitian generators  $X_{k}$ of the
fundamental representation of the $SU(3)$ group are defined by
the Gell-Mann matrices
\begin{equation}\label{00000001x}
\begin{array}{l}
\displaystyle {X}_{1}={\phantom{\ii}}\left(e^{(1)}\otimes e^{(2)}\right)+{\phantom{\ii}}\left(e^{(2)}\otimes e^{(1)}\right), \\[5pt]
\displaystyle {X}_{2}=\ii\left(e^{(2)}\otimes e^{(1)}\right)-\ii\left(e^{(1)}\otimes e^{(2)}\right), \\[5pt]
\displaystyle {X}_{3}={\phantom{\ii}}\left(e^{(1)}\otimes e^{(1)}\right)-{\phantom{\ii}}\left(e^{(2)}\otimes e^{(2)}\right), \\[5pt]
\displaystyle {X}_{4}={\phantom{\ii}}\left(e^{(1)}\otimes e^{(3)}\right)+{\phantom{\ii}}\left(e^{(3)}\otimes e^{(1)}\right), \\[5pt]
\displaystyle {X}_{5}=\ii\left(e^{(3)}\otimes e^{(1)}\right)-\ii\left(e^{(1)}\otimes e^{(3)}\right), \\[5pt]
\displaystyle {X}_{6}={\phantom{\ii}}\left(e^{(2)}\otimes e^{(3)}\right)+{\phantom{\ii}}\left(e^{(3)}\otimes e^{(2)}\right), \\[5pt]
\displaystyle {X}_{7}=\ii\left(e^{(3)}\otimes e^{(2)}\right)-\ii\left(e^{(2)}\otimes e^{(3)}\right), \\[5pt]
\displaystyle {X}_{8}={\phantom{\ii}}\left(e^{(1)}\otimes e^{(1)}\right)+{\phantom{\ii}}\left(e^{(2)}\otimes e^{(2)}\right)\!-2\left(e^{(3)}\otimes e^{(3)}\right).
\end{array}
\end{equation}
\noindent The irreducibility condition for this representation is the matrix Dirac equation (see, e.g., \cite{Barut})
\begin{equation}\label{0000001}
\left( \ii\gamma^\mu\partial_\mu{\mathds{I}} -m{\mathds{I}}\right)\Psi
(x)=0.
\end{equation}
\noindent Obviously, in this  representation all
the  components of the  multiplet have equal masses.

Let us turn to another  representation of the extended
symmetry group of the theory for each multiplet. The idea of the
proposed transformation is based on the fact that the derivation
algebra of the Poincar\'{e} algebra contains not only the
operators $P^{\mu}, M^{\mu\nu}$, but also the generator of
dilatation $D$ (see, e.g., \cite{Barut})
\begin{equation}\label{dil}
\left[P^{\mu}, D \right]=P^{\mu}, \quad \left[M^{\mu\nu}, D \right]=0.
\end{equation}
\noindent Because of this, it is possible to construct an outer
automorphism of the direct product of the Poincar\'{e} group and
$SU(3)$. For the  representation in the space $ {\cal{H}}_{m
,1/2}$ the automorphism leads to  scaling transformations of
the coordinates. The transformations can vary for different
subspaces. As a consequence, it will allow  to consider
multiplets with different masses of the components.

To be specific, we assume that we work with the  neutrino
multiplet. Introduce  new basis vectors $n^{(l)}$ with the help of a
unitary matrix ${\mathds{V}}^{(\nu)}$ acting on  $e^{(l)}$
\begin{equation}\label{000000001a}
n^{(l)}_{s}=\sum_{r=1}^{3}V_{sr}^{(\nu)}e^{(l)}_{r}.
\end{equation}
\noindent The new representation space $
{{{\cal{H}}}}_{m,1/2}^{(\nu)}$  will also be  defined as the
direct sum of the  spaces of the Dirac representations  of
the Poincar\'{e} group
\begin{equation}\label{01x}
  {{{\cal{H}}}}_{m,1/2}^{(\nu)}=
  {{\cal{H}}}_{m,1/2}^{(\nu_1)}
  \oplus{{\cal{H}}}_{m,1/2}^{(\nu_2)}
  \oplus{{\cal{H}}}_{m,1/2}^{(\nu_3)}.
\end{equation}
\noindent We define a basis in ${{\cal{H}}}_{m,1/2}^{(\nu)}$
in the form
\begin{equation}\label{001x}
  {\Psi}^{(\nu)}_{q,\zeta,{\kern 0.6pt }\mu_{\kern 0.4pt l}}(x)= \psi^{(\nu)}_{q,\zeta,{\kern 0.6pt }\mu_{\kern 0.4pt l}}(x)n^{(l)},
\end{equation}
\noindent where $\psi^{(\nu)}_{q,\zeta,{\kern 0.6pt }\mu_{\kern
0.4pt l}}(x)$ are the  plane waves derived from \eqref{0001} by
a dilatation of the   coordinates
\begin{equation}\label{0001x}
  \psi_{q,\zeta,{\kern 0.6pt }\mu_{\kern 0.4pt l}}^{(\nu)}(x)=\frac{\mu_{l}^{3/2}}{\sqrt{2q^{0}}}
  u_{q,\zeta}e^{-\ii \mu_{l}(qx)}.
\end{equation}
\noindent Here $q^{0}=\sqrt{{\bf q}^{2}+m^{2}}$, and $\mu_{l}$ are positive real numbers. The spinors $u_{q,\zeta}$ satisfy the equation
\begin{equation}\label{00010x}
(\gamma^{\mu}q_{\mu}-m)u_{q,\zeta}=0,
\end{equation}
\noindent  and are normalized by the condition
\begin{equation}\label{000100d}
{\bar{u}}_{q,\zeta}{u}_{q,\zeta'}=
2m\delta_{\zeta,\zeta'}.
\end{equation}

Evidently, ${\Psi}^{(\nu)}(x)$ (the elements of the
space ${{{\cal{H}}}}_{m,1/2}^{(\nu)}$) and ${\Psi}(x)$ (the
elements of the  space  ${{\cal{H}}}_{m,1/2}$) are connected
by the unitary (with respect  to  the  scalar product
\eqref{000001}) transformation
\begin{equation}\label{00020x}
{\Psi}^{(\nu)}(x)={\cal K}^{(\nu)}{\Psi}(x) = \int K(x,y)\,{\Psi}(y)\, d{\bf y}.
\end{equation}
\noindent Its kernel  is determined by the formula
\begin{equation}\label{000020x}
K(x,y)=\frac{1}{(2\pi)^{3}}\sum\limits_{l=1}^{3}
\sum\limits_{\zeta=\pm 1}\int d{\bf q}\,{\Psi}^{(\nu)}_{q,\zeta,{\kern 0.6pt }
\mu_{\kern 0.4pt l}}(x)\otimes
{\Psi}^{\dag}_{q,\zeta,l}(y).
\end{equation}
\noindent The explicit form of  the dilatation  operator is
\begin{equation}\label{vrio}
D=x_{\alpha}\partial^{\alpha}+3/2.
\end{equation}
\noindent Therefore, the intertwining operator  for the representations in the spaces
${{{\cal{H}}}}_{m,1/2}^{(\nu)}$ and ${{{\cal{H}}}}_{m,1/2}$ realizing the transformation  \eqref{00020x} is
\begin{equation}\label{aaa0}
{\cal K}^{(\nu)} =\sum\limits_{l=1}^{3}
{\cal D}_{(l)}\left(n^{(l)}\otimes e^{(l)}\right),
\end{equation}
\noindent where
\begin{equation}\label{aaa1}
{\cal D}_{(l)}=\exp{\big(\ln\mu_{l}(x_{\alpha}\partial^{\alpha}+3/2)\big)}.
\end{equation}

A direct calculation yields  the explicit form  of the elements of
the Lie algebra of the considered representation. The  action of
the generators ${X}_{k}^{(\nu)}={\cal
K}^{(\nu)}{X}_{k}{\cal K}^{(\nu)-1}$ of the  $ SU (3) $ group on
the basis \eqref{001x} is reduced to  obvious permutations of its
elements
\begin{equation}\label{0000200x}
{X}_{1}^{(\nu)}{\Psi}^{(\nu)}_{q,\zeta,{\kern 0.6pt }\mu_{\kern 0.4pt 1}}(x)=
{\Psi}^{(\nu)}_{q,\zeta,{\kern 0.6pt }\mu_{\kern 0.4pt 2}}(x),\;\; {X}_{1}^{(\nu)}{\Psi}^{(\nu)}_{q,\zeta,{\kern 0.6pt }\mu_{\kern 0.4pt 2}}(x)=
{\Psi}^{(\nu)}_{q,\zeta,{\kern 0.6pt }\mu_{\kern 0.4pt 1}}(x),\;\; {X}_{1}^{(\nu)}{\Psi}^{(\nu)}_{q,\zeta,{\kern 0.6pt }\mu_{\kern 0.4pt 3}}(x)=0,
\end{equation}
\noindent  and so on.   The diagonal operators
${X}_{3}^{(\nu)},{X}_{8}^{(\nu)}$ are determined by numerical
matrices. The explicit form of these operators see in 
\ref{A.0}.

The generators of the Lorentz group do not change the form
\begin{equation}\label{000000001}
{M}_{\mu\nu}^{(\nu)}
={\cal K}^{(\nu)}M_{\mu\nu}{\cal K}^{(\nu)-1}
=\ii\Big((x_{\mu}\partial_{\,\nu}-x_{\nu}
\partial_{\,\mu})+
(\gamma_{\mu}\gamma_{\nu}-\gamma_{\nu}
\gamma_{\mu})/4\Big)
{\mathds{I}}\,,
\end{equation}
\noindent since the dilatation operator commute with the
generators of rotations and boosts (see \eqref{dil}). This fact is
quite natural, because the results of observations, in particular,
of the study of particle oscillations cannot depend on the choice
of the inertial reference frame used for measurement.

For the translation generators, we have
\begin{equation}\label{000000001y}
 {P}_{\mu}^{(\nu)}={\cal K}^{(\nu)}{P}_{\mu}{\cal K}^{(\nu)-1} = \ii{\partial_{\,\mu}}{\mathds{N}}^{(\nu)},
\end{equation}
\noindent where
\begin{equation}\label{0000000001}
{\mathds{N}}^{(\nu)}  = \sum\limits_{l=1}^{3}\frac{1}{\mu_{l}}
    \left(n^{(l)}\otimes \na{l}{\phantom{j}}\right).
\end{equation}
\noindent That is $\mu_{l}^{-1}$ are  the eigenvalues of  the
matrix ${\mathds{N}}^{(\nu)}$,   and $ n^{(l)} $ are its
eigenvectors normalized by the conditions (the asterisk denotes  the
complex conjugation)
\begin{equation}\label{3}
\sum\limits_{s=1}^{3}n^{(l)}_{s}\na {k}{s}=\delta_{kl},\qquad
\sum\limits_{l=1}^{3}n^{(l)}_{s}\na {l}{r}=\delta_{sr}.
\end{equation}
\noindent The operators
\begin{equation}\label{2}
    {\mathds{P}}^{(\nu)}_{(l)}=n^{(l)}\otimes \na{l}{\phantom{j}}
\end{equation}
\noindent are orthogonal projectors
\begin{equation}\label{01z}
{\mathds{P}}^{(\nu)}_{(l)}{\mathds{P}}^{(\nu)}_{(k)}=
\delta_{kl}{\mathds{P}}^{(\nu)}_{(l)},\quad \sum\limits_{l=1,2,3} {\mathds{P}}^{(\nu)}_{(l)}={\mathds{I}}.
\end{equation}

The Dirac equation, i.e. the irreducibility condition for this representation,
is now written as
\begin{equation}\label{4}
\left( \ii\gamma^\mu\partial_\mu{\mathds{N}}^{(\nu)} -m{\mathds{I}}\right)
{\Psi}^{(\nu)}(x)=0.
\end{equation}
\noindent  Since the parameters $ \mu_{l} $ are chosen nonzero
then $ {\mathds{N}}^{(\nu)} $ is non-degenerate and the inverse
matrix exists
\begin{equation}\label{5}
({\mathds{N}}^{(\nu)})^{-1}=\sum\limits_{l=1}^{3} {\mu_{l}}
    {\mathds{P}}^{(\nu)}_{(l)}.
\end{equation}
\noindent If we multiply \eqref{4} by \eqref{5} then the Dirac
equation takes a more familiar form
\begin{equation}\label{6}
\left( \ii\gamma^\mu\partial_\mu{\mathds{I}}  -{\mathds{M}}^{(\nu)}\right)
{\Psi}^{(\nu)}(x)=0,
\end{equation}
\noindent  where
\begin{equation}\label{7}
{\mathds{M}}^{(\nu)}=m({\mathds{N}}^{(\nu)})^{-1}=
\sum\limits_{l=1}^{3}{m_{l}}{\mathds{P}}_{(l)}^{(\nu)},
\quad m_{l}={\mu_{l}}m.
\end{equation}
\noindent The sets of solutions  of equations \eqref{4} and
\eqref{6} coincide. Further, we  will refer only to  equation
\eqref{6}.

We emphasize once again that the eigenvalues of the Casimir
operators  constructed from the generators \eqref{000000001},
\eqref{000000001y}, and \eqref{aaa2} take  the same values on the
solutions of these equations  and on  the  elements of the
initial representation  space ${{\cal{H}}}_{m,1/2}$.  In
particular,
${{P}^{(\nu)}}^{\mu}{P}_{\mu}^{(\nu)}{\Psi}^{(\nu)}(x)
=m^{2}{\Psi}^{(\nu)}(x)$, but now the parameter $m$ is not the
observed mass, it rather sets the scale of the multiplet masses.
In other words, it represents the bare mass of the multiplet. The
observed masses  are determined by the action of the canonical
momentum operator squared on the functions
${{\Psi}^{(\nu)}_{q,\zeta,{\kern 0.6pt } \mu_{\kern 0.4pt
l}}(x)}$ and are equal to $m_{l}=\mu_{l}m$ for  the $l$-th
component of the multiplet.

Let us discuss this point in detail. The general solution of
\eqref {6} can be expanded in the basis  functions that are the
eigenfunctions of a complete set of operators commuting with the
operator of the equation. Naturally, the choice of the complete
set is ambiguous. For the standard  Dirac equation, if we restrict
ourselves to the case of plane-wave solutions,  the complete set
includes the space components of the momentum (i. e. the
generators of the space translations) and the operator of the spin
projection. This operator  is constructed in  the common way from
the components of the Pauli--Lubanski--Bargmann vector. The
ambiguity of the  complete set composition  in that case depends
on  the choice of the spin operator{\footnote{This is true  only
for the Dirac equation describing free particles. The ambiguity of
the choice of the operators with continuous spectrum arises, for
example, in the problem  of  the neutrino spin  evolution
for a neutrino propagating in a dense matter and  an
electromagnetic field \cite{arlomur}.}}. In contrast to the
standard Dirac equation,  the complete set for Eq. \eqref{6}
contains five operators. This circumstance makes the situation
more complicated.

We now find the complete sets of operators for which the basis
of  the solution  space  of Eq. \eqref{6} consists of plane
waves. Let ${\mathds{L}}^{(\nu)}$  be a non-degenerate Hermitian
matrix which commutes with  ${{\mathds{M}}}^{(\nu)}$. This matrix
can be written as
\begin{equation}\label{07}
{\mathds{L}}^{(\nu)}=
\sum\limits_{l=1}^{3}{\lambda_{l}}
{\mathds{P}}_{(l)}^{(\nu)},\quad {\lambda_{l}}\neq 0.
\end{equation}
\noindent If we assume that the parameters ${\lambda_{l}}$ are
pairwise  distinct,  then, in the space of the solutions of Eq.
\eqref{6},  this matrix defines  an action of the operator
${\mathcal{M}}^{(\nu)} $, which is a linear combina\-tion of
operators ${X}_{3}^{(\nu)}$, ${X}_{8}^{(\nu)}$, and the Casimir
operators of $SU(3)$.  Since the representation is irreducible,
then the action of the Casimir operators is given  by  the
matrices that are multiple of the identity matrix. Obviously,
${\mathcal{M}}^{(\nu)}$  can be chosen as one of the operators of
the complete set. This operator is used to isolate the orthogonal
subspaces determined by the projectors
${\mathds{P}}_{(l)}^{(\nu)}$. Therefore, the numerical values of
the parameters ${\lambda_{l}}$ can be arbitrary. It is natural to
set $ {\lambda_{l}} = m_{l} $, that is, to assume that the action
of the operator ${\mathcal{M}}^{(\nu)}$ is defined by the matrix
${{\mathds{M}}}^{(\nu)}$.

For a basis to be a plane-wave one, the complete set must
include three operators with continuous spectrum. If we choose
the space translation generators, that is, the spatial
components of the 4-vector ${P}_{\mu}^{(\nu)}$, as such operators,
then any standard spin projector multiplied by
${{\mathds{N}}}^{(\nu)} $ will determine the spin projection. In
this case the complete orthonormal system of solutions of
Eq. \eqref{6} corresponding to  a fixed (positive) frequency
coincides with the  basis ${\Psi}^{(\nu)}_{q,\zeta,{\kern 0.6pt
}\mu_{\kern 0.4pt l}}(x)$ (see Eq. \eqref{001x}), which was considered
previously.

However, we can take the spatial components of the 4-vector
\begin{equation}\label{016}
 {\mathds{L}}^{(\nu)}{P}_{\mu}^{(\nu)}= \ii\partial_{\mu}{\mathds{L}}^{(\nu)}
 {\mathds{N}}^{(\nu)}
\end{equation}
\noindent to play the  role of the operators with continuous
spectrum, where ${\mathds{L}}^{(\nu)}$ is any matrix which
satisfies \eqref{07}. In particular, we can  set
${{\mathds{L}}}^{(\nu)}=({{\mathds{N}}}^{(\nu)})^{-1}$. Then the
basis functions are the eigenfunctions of the spatial
components of the canonical momentum operator $i\partial_{\mu}$,
and
\begin{equation}\label{001xx}
  {\Psi}^{(\nu)}_{p,\zeta,m_{l}}(x)= \psi_{p,\zeta,m_{l}}^{(\nu)}(x)n^{(l)},
\end{equation}
\noindent where $\psi_{p,\zeta,m_{l}}^{(\nu)}(x)$ are the plane
waves describing particles with masses $m_{l}$
\begin{equation}\label{0001xx}
  \psi_{p,\zeta,m_{l}}^{(\nu)}(x)=\frac{1}
  {\sqrt{2p^{0}_{l}}}\,
  u_{p,\zeta}^{(m_{l})}e^{-\ii (px)}.
\end{equation}
\noindent Here $p^{0}_{l}=\sqrt{{\bf p}^{2}+m^{2}_{l}}$, the
spinors $u_{p,\zeta}^{(m_{l})}$ satisfy the equation
\begin{equation}\label{00010xx}
(\gamma^{\mu}p_{\mu}-m_{l})u_{p,\zeta}^{(m_{l})}=0,
\end{equation}
\noindent  and are normalized by the condition
\begin{equation}\label{000100xx}
{\bar{u}}_{p,\zeta}^{(m_{l})}
{u}_{p,\zeta'}^{(m_{l})}=2m_{l}\delta_{\zeta,\zeta'}.
\end{equation}

Thus, the matrix ${\mathds{M}}^{(\nu)}$ can be interpreted as
the mass matrix of the neutrino multiplet, and  the
parameters $m_{l}= \mu_{l} m$ are  the observed masses of the
particles. The states that are described by the eigenfunctions of
${\mathcal{M}}^{(\nu)}$, can be naturally  called the mass
states for any choice of the operators with continuous spectrum,
that is, for any choice of the matrix ${\mathds{L}}^{(\nu)}$
including the case ${\mathds{L}}^{(\nu)}={\mathds{I}}$.

However, we can choose a basis  in the space $
{{{\cal{H}}}}^{(\nu)}_{m,1/2}$   in the form of a superposition of
the mass states. Consider an arbitrary unitary matrix
${{\mathds{U}}}$ with components $ U_{\alpha l} $. The
functions
\begin{equation}\label{016x}
{\Psi}_{p,\zeta,\alpha}^{(\nu)}(x)=
\sum\limits_{l=1}^{3}
U_{\alpha l}{\Psi}^{(\nu)}_{p,\zeta,l}(x),
\end{equation}
\noindent where $ {\Psi}_{p,\zeta, l}^{(\nu)}(x) $ is the wave
function of an arbitrary mass state,  make up a complete
orthonor\-mal set in ${\cal{H}}^{(\nu)}_{m,1/2}$ with the
scalar product \eqref{000001}. The basis \eqref{016x} can be
obtained as a result of the unitary transformation ${{\cal{U}}}$
of the space ${{{\cal{H}}}}^{(\nu)}_{m,1/2}$ onto itself
\begin{equation}\label{00020xx}
{\Psi}_{p,\zeta,\alpha}^{(\nu)}(x)={{\cal{U}}}\,
{\Psi}_{p,\zeta,l}^{(\nu)}(x)
=\int \widetilde{K}(x,y)\,{\Psi}_{p,\zeta,l}^{(\nu)}(y)\, d{\bf y}.
\end{equation}
\noindent The kernel of the transformation  is determined by the formula
\begin{equation}\label{000020xx}
\widetilde{K}(x,y)=\frac{1}{(2\pi)^{3}}
\sum\limits_{\alpha=1}^{3}
\sum\limits_{k=1}^{3}\sum\limits_{\zeta=\pm 1}\int d{\bf p}\,\delta_{\alpha k}{\Psi}_{p,\zeta,\alpha}^{(\nu)}(x)\otimes
{{\Psi}_{p,\zeta,k}^{(\nu)\dag}}(y).
\end{equation}

The elements of the basis \eqref{016x} are no longer
eigenfunctions of the operator  ${{\cal{M}}}^{(\nu)}$ determined by
the mass matrix. The fifth operator of the complete set (denote
it as ${{\cal{F}}}^{(\nu)}$) is defined now as
\begin{equation}\label{000020xxx}
{{\cal{F}}}^{(\nu)}= {\cal{U}}{\cal{M}}^{(\nu)}{\cal{U}}^{-1}.
\end{equation}
\noindent  This operator can explicitly depend on  the event
space coordinates. It should be emphasized that the form of the
causal Green function for Eq. \eqref{6} does not depend on the
chosen basis
\begin{equation}\label{com27}
S_{c}^{(\nu)}(x)=\frac{1}{(2\pi)^4}\sum\limits_{l=1}^{3}\,
{\mathds
P}^{(\nu)}_{l}\!\!\int\frac{(\gamma_{\mu}{p}^{\mu} +m_{l})\,
e^{-\ii(px)}}{m^2_{l}-p^2-\ii\epsilon}\,d^4p.
\end{equation}

Similarly, we can construct the representation spaces for the
multiplets of the charged leptons  and  the down- and up-type
quarks, where the values $ \mu_{l} $, and the matrices ${\mathds
{V}}, {\mathds{N}}, {\mathds{M}}, {\mathds{L}} $ for these
multiplets may be different.
Since we apply this  approach to the problem of neutrino
oscillations  as an example, we introduce special notations for
the multiplet of the charged leptons. The  bare mass is denoted
by $ M $, without assuming in advance that $ M = m $. According
to the  tradition in the theory of oscillations we denote the
subspaces of  the mass states by Greek letters.  We will
write the dilatation parameters as $ \eta_{\beta} $ instead of $
\mu_{l} $. Accordingly, the physical masses of the charged
leptons  will be $M_{\beta}=\eta_{\beta}M, \;\; \beta = e, \mu,
\tau$.

Applying a unitary matrix ${\mathds{V}}^{(e)}$ to $e^{(l)}$,
we obtain  new basis vectors $a^{(\beta)} $
\begin{equation}\label{000000001f}
\displaystyle a^{(\beta)}_{s}=\sum\limits_{r=1}^{3}V_{sr}^{(e)}e^{(l)}_{r},
\;\;\;
\sum\limits_{s=1}^{3}a^{(\alpha)}_{s}\ap{\beta}{s} =
\delta_{\alpha\beta}, \;
\sum\limits_{\beta=e,\mu,\tau}\!\! a^{(\beta)}_{s}\ap {\beta}{r}=\delta_{sr}.
\end{equation}
\noindent Therefore, the orthogonal projectors for the representation space of the charged leptons take the form
\begin{equation}\label{2a}
    {\mathds{P}}^{(e)}_{(\beta)}=a^{(\beta)}\!\otimes \ap{\beta}{\phantom{j}}.
\end{equation}
\noindent Hence,
\begin{equation}\label{10}
    {\mathds{N}}^{(e)}=\sum\limits_{\beta=e,\mu,\tau}
    \frac{1}
    {{{\eta}}_{\beta}}\,
    {\mathds{P}}^{(e)}_{(\beta)},\quad {\mathds{M}}^{(e)}=\sum\limits_{\beta=e,\mu,\tau}
    M_{\beta}
    {\mathds{P}}^{(e)}_{(\beta)}.
\end{equation}
\noindent We determine  the basis elements of
${{\cal{H}}}_{m,1/2}^{(e)}$,  corresponding to the basis
\eqref{001x}, as
\begin{equation}\label{001xxxxx}
  {\Psi}^{(e)}_{q,\zeta,{\kern 0.6pt }\eta_{\kern 0.4pt \beta}}(x)= \psi^{(e)}_{q,\zeta,{\kern 0.6pt }\eta_{\kern 0.4pt \beta}}(x)\,a^{(\beta)},
\end{equation}
\noindent where
\begin{equation}\label{0001xxxx}
  \psi_{q,\zeta,{\kern 0.6pt }\eta_{\kern 0.4pt \beta}}^{(e)}(x)=\frac{\eta_{\beta}^{3/2}}{\sqrt{2q^{0}}}
  u_{q,\zeta}e^{-\ii \eta_{\beta}(qx)}.
\end{equation}
\noindent We determine  the basis elements of
${{\cal{H}}}_{m,1/2}^{(e)}$, corresponding to  the basis
\eqref{001xx}, as
\begin{equation}\label{001xxx}
  {\Psi}^{(e)}_{p,\zeta,M_{\beta}}(x)= \psi_{p,\zeta,M_{\beta}}^{(e)}(x)\,a^{(\beta)},
\end{equation}
\noindent where $\psi_{p,\zeta,M_{\beta}}^{(e)}(x)$ are the
plane waves describing particles with mass $M_{\beta}$
\begin{equation}\label{0001xxx}
  \psi_{p,\zeta,M_{\beta}}^{(e)}(x)=
  \frac{1}{\sqrt{2p^{0}_{l}}}\,
  u_{p,\zeta}^{(M_{\beta})}e^{-\ii (px)}.
\end{equation}

We  do not introduce special notations for  the quark multiplets,
but use  the appropriate indices only.

\section{Modified model of electroweak interactions}\label{M.2}

Now we are able to write the  Lagrangian of the modified theory
describing the electroweak interactions. If, as usual, the
interaction is carried  by the gauge fields associated with the
group $SU(2)\times U(1) $, then the Lagrangian of such a theory is
the Lagrangian of the minimally extended Standard Model with  changes in the fermion sector only.

The Lagrangian for the physical fermion fields in our model is written as follows:
\begin{equation}\label{l2}
{\cal{L}}_{f}={\cal{L}}_{0}+{\cal{L}}_{int},
\end{equation}
\noindent where
\begin{equation}\label{l3}
\displaystyle {\cal{L}}_{0}= \frac{\ii}{2}\sum\limits_{i=\nu, e,u,d}\left[\left(\bar{\Psi}^{(i)}
\gamma^{\mu}
(\partial_{\mu}{\Psi}^{(i)})\right)
-(\partial_{\mu}\bar{\Psi}^{(i)})
\gamma^{\mu}{\Psi}^{(i)}
   \right]\;-\!\!\!\!\!\sum\limits_{i=\nu, e,u,d}\!\!\!
   \bar{\Psi}^{(i)}{\mathds{M}}^{(i)}{\Psi}^{(i)}
\end{equation}
\noindent is the Lagrangian of free fields and
\begin{equation}\label{l1}
\begin{array}{c}
\displaystyle {\cal{L}}_{int}= -\!\!\!\!\!\sum\limits_{i=\nu, e,u,d}\!\!\!\bar{\Psi}^{(i)}
   {\mathds{M}}^{(i)}({H}/{v}){\Psi}^{(i)}\\
\displaystyle -\frac{g}{2\sqrt{2}}
\left(\bar{\Psi}^{(e)}\gamma^{\mu}
(1+\gamma^{5})\,{\Psi}^{(\nu)}
W_{\mu}^{-}+\bar{\Psi}^{(\nu)}\gamma^{\mu}
(1+\gamma^{5})\,{\Psi}^{(e)}
W_{\mu}^{+}\right)\\[12pt]
\displaystyle -\frac{g}{2\sqrt{2}}
\left(\bar{\Psi}^{(d)}\gamma^{\mu}
(1+\gamma^{5})\,{\Psi}^{(u)}
W_{\mu}^{-}+\bar{\Psi}^{(u)}\gamma^{\mu}
(1+\gamma^{5})\,{\Psi}^{(d)}
W_{\mu}^{+}\right)\\
\displaystyle -e\!\!\!\!\!\sum\limits_{i= e,u,d}Q^{(i)}\bar{\Psi}^{(i)}\gamma^{\mu}\,
  {\Psi}^{(i)}A_{\mu}\\
\displaystyle -\frac{g}{2\cos\theta_{\mathrm{W}}}\!\sum\limits_{i=\nu, e,u,d}\!\!
\bar{\Psi}^{(i)}\gamma^{\mu}
\left(T^{(i)}-2Q^{(i)}\sin^{2}
\theta_{\mathrm{W}}
+T^{(i)}\gamma^{5}\right)
{\Psi}^{(i)}Z_{\mu}
\end{array}
\end{equation}
\noindent is the Lagrangian  of the interaction between the
fermion fields, the vector boson fields $W^{\pm}_{\mu},
Z_{\mu}, A_{\mu}$, and the Higgs field $H$. Here
$\theta_{\mathrm{W}}$ is the Weinberg angle,
$e=g\sin\theta_{\mathrm{W}}$ is the positron electric charge,
$T^{(i)}$ is the weak isospin projection ($T^{(\nu)}=T^{(u)}= 1/2,\,
T^{(e)}=T^{(d)}= -1/2$), $Q^{(i)}$ is the electric charge of the
multiplet in the units of $e$, and $v$ is the vacuum
expectation value of the Higgs field.
Thus, \eqref{l2} formally coincides with the Lagrangian of the
Standard Model, supplemented with the singlets of the
right-handed neutrinos (see, e.g., \cite{pdg}).
However,  the field
functions $ {\Psi}^{(i)}$ describe not the individual
particles, but the multiplet as a whole. So it is not
necessary to introduce the mixing matrices into
${\cal{L}}_{int}$ explicitly.

The action defined by the Lagrangian of free fields \eqref{l3} is
explicitly  invariant with respect to $SU(3)$ transformations generated by $X_{k}^{(\nu)}, {X}_{k}^{(e)}, {X}_{k}^{(u)},$ and ${X}_{k}^{(d)} $ (see \eqref{0000200x}). Therefore,  when quantizing
the model, the multiplet can be considered as a single
particle. The one-particle states in the
Fock space are defined as usual, the creation and annihilation
operators satisfy the canonical commutation relations. However,
these operators carry an additional discrete quantum number that is
associated with the mass  of the state. The
multiplet can  be either in one of the three mass states, or in a
pure quantum state that is a superposition of the states with
fixed masses. In a certain sense we may
 say that there are only four fundamental fermions in our
model.

We will discuss all  this in detail, using the approach
described in \cite{BSh}. Consider the basis \eqref{001x}. We can write
the components of the field functions as
\begin{equation}\label{s1}
\begin{array}{c}
\displaystyle   {\Psi}^{(\nu)}_{s}(x)= \frac{1}{(2\pi)^{3/2}}\sum\limits_{l=1}^{3}
\sum\limits_{\zeta=\pm 1}\int \frac{d{\bf q}}{\sqrt{2q^{0}}}\;{\mu_{l}^{3/2}}n^{(l)}_{s}\\[7pt] \displaystyle \times\left[
  e^{-\ii \mu_{l}(qx)}u_{q,\zeta}{\sf a}^{-}_{l,\zeta}({\bf q})+
  e^{\ii \mu_{l}(qx)}u_{-q,\zeta}{\sf a}^{+}_{l,\zeta}({\bf q})\right],\\[7pt]
\displaystyle   \bar{{\Psi}}^{(\nu)}_{s}(x)= \frac{1}{(2\pi)^{3/2}}\sum\limits_{l=1}^{3}
\sum\limits_{\zeta=\pm 1}\int \frac{d{\bf q}}{\sqrt{2q^{0}}}\;{\mu_{l}^{3/2}}\na{l}{s}\\[7pt] \displaystyle \times\left[
  e^{-\ii \mu_{l}(qx)}\bar{u}_{-q,\zeta}\as{-}{l,\zeta}({\bf q})+
  e^{\ii \mu_{l}(qx)}\bar{u}_{q,\zeta}\as{+}{l,\zeta}({\bf q})\right].
\end{array}
\end{equation}

With the help of Noether's theorem we can find the integrals of
motion for the fields.   An analog of the energy-momentum
tensor associated with the translational symmetry is defined by
the relation
\begin{equation}\label{s2}
 T_{\alpha\beta}=\frac{1}{2}\sum\limits_{s=1}^{3}
 \Big[\bar{{\Psi}}^{(\nu)}_{s}(x)\gamma_{\alpha}
 \big({P}_{\beta}^{(\nu)}
 {\Psi}^{(\nu)}(x)\big)_{s}+
 \big({P}_{\beta}^{(\nu)*}
 \bar{\Psi}^{(\nu)}(x)\big)_{s}\gamma_{\alpha}
 {\Psi}^{(\nu)}_{s}(x)\Big].
\end{equation}
\noindent It ensures the existence of the conserved  vector
\begin{equation}\label{s5}
    {\mathcal{P}}_{\beta}=\sum\limits_{l=1}^{3}
    \sum\limits_{\zeta=\pm 1}\int q_{\beta}{d{\bf q}}\Big[\as{+}{l,\zeta}({\bf q}){\sf{a}}^{-}_{l,\zeta}({\bf q})-\as{-}{l,\zeta}({\bf q}){\sf a}^{+}_{l,\zeta}({\bf q})\Big].
\end{equation}
\noindent The current vector associated with the global gauge
symmetry of the Lagrangian is defined by the relation
\begin{equation}\label{s3}
   J_{\alpha}=\sum\limits_{s=1}^{3}
   \bar{\Psi}^{(\nu)}_{s}(x)\gamma_{\alpha}
   {\Psi}^{(\nu)}_{s}(x).
\end{equation}
\noindent It ensures the conservation of the total field
charge
\begin{equation}\label{s6}
    {\mathcal{Q}}=\sum\limits_{l=1}^{3}
    \sum\limits_{\zeta=\pm 1}\int {d{\bf q}}\Big[\as{+}{l,\zeta}({\bf q}){\sf a}^{-}_{l,\zeta}({\bf q})+\as{-}{l,\zeta}({\bf q}){\sf a}^{+}_{l,\zeta}({\bf q})\Big].
\end{equation}
\noindent The tensor associated with the $SU(3)$-symmetry of
the fields is defined  by the relation
\begin{equation}\label{s4}
S_{k\alpha}  = \frac{1}{2}\sum\limits_{s=1}^{3}
\Big[\bar{\Psi}^{(\nu)}_{s}(x)
\gamma_{\alpha}\big({X}_{k}^{(\nu)}
{\Psi}^{(\nu)}(x)\big)_{s}\\
+\big({X}_{k}^{(\nu)*}
\bar{\Psi}^{(\nu)}(x)\big)_{s}
\gamma_{\alpha}{\Psi}^{(\nu)}_{s}(x)
\Big],\;\; k=1...8.
\end{equation}
\noindent It ensures the  existence of eight integrals of
motion (see \eqref{s7} -- \eqref{s14}). Using these integrals of motion and the total field
charge it is possible to construct nine linear combinations
\begin{equation}\label{s15}
    {\mathcal{X}}_{lk}=\sum\limits_{\zeta=\pm 1}\int {d{\bf q}}\Big[\as{+}{l,\zeta}({\bf q}){\sf a}^{-}_{k,\zeta}({\bf q})+\as{-}{l,\zeta}({\bf q}){\sf a}^{+}_{k,\zeta}({\bf q})\Big],\quad k,l=1,2,3,
\end{equation}
\noindent and three of them are diagonal in the indices
$l,k$.

We do not consider the angular momentum tensor,
because its analysis is not critical for what follows. Since the
Lorentz group generators \eqref{000000001} have the standard
form, then the angular momentum tensor is the same as in
\cite{BSh}.

The integrals of motion for the fields are associated with the
generators of the Lie algebra of the symmetry group of the theory
as follows:
\begin{equation}\label{s16}
 Z{\Psi}^{(\nu)}(x)=
 \left[{\Psi}^{(\nu)}(x),{\cal{Z}}\right].
\end{equation}
\noindent Here $Z=\left\{P^{(\nu)}_{\beta}, Q, X^{(\nu)}_{k}\!\!\!...\right\}$ and
${\cal{Z}}=\left\{{\cal{P}}_{\beta}, {\cal Q}, {\cal X}_{k}...\kern-0.4pt\right\}$. These relations enable one to interpret
$ \as{+}{l,\zeta}({\bf q})$ and $ {\sf
a}^{-}_{l,\zeta}({\bf q}) $  as the operators of neutrino
creation and annihilation  in the state ${l}$ with the kinetic
momentum $ {\bf q} $ and the polarization $ \zeta$. Accordingly, $
{\sf a}^{+}_{l,\zeta}({\bf q})$ and $\as{-}{l,\zeta} ({\bf q})$
are the operators of antineutrino creation and annihilation  in
the state ${l}$ with the kinetic momentum $ {\bf q} $ and  the
polarization $\zeta $. Eq. \eqref{s16} and the invariance condition  with respect to  the change
of particles  to antiparticles for $ {\cal{P}}_{\beta} $  yield the canonical commutation relations
\begin{equation}\label{s17}
\begin{array}{l}
\Big[\,{\sf a}^{-}_{l,\zeta}({\bf q}), \as{+}{k,\zeta'}({\bf q}')\Big]_{+}\!\!=\delta_{lk}\delta_{\zeta \zeta'}\delta({\bf q}-{\bf q}'), \\
\Big[\,\as{-}{l,\zeta}({\bf q}),{\sf a}^{+}_{k,\zeta'}({\bf q}')\Big]_{+}\!\!=\delta_{lk}
\delta_{\zeta \zeta'}\delta({\bf q}-{\bf q}').
\end{array}
\end{equation}

Let us now consider linear combinations of these operators
\begin{equation}\label{s18}
{\sf a}^{\pm}_{\alpha,\zeta}({\bf q})=\sum\limits_{l=1}^{3}U_{\alpha l}{\sf a}^{\pm}_{l,\zeta}({\bf q}),\;\; \as{\pm}{\alpha,\zeta}({\bf q})=\sum\limits_{l=1}^{3}U_{\alpha l}^{*}\as{\pm}{l,\zeta}({\bf q}),
\end{equation}
\noindent where $ U_{\alpha l} $ are components of an arbitrary unitary matrix ${{\mathds{U}}}$.  The commutation relations   for ${\sf
a}^{\pm}_{\alpha,\zeta}({\bf q}), \as{\pm}{\alpha,\zeta}({\bf q})$
are canonical. When expressed in terms of these
operators, only   ${\cal P}_{\beta}$ and ${\cal Q}$ are diagonal.
However, using $ {\cal Q} $ and the remaining integrals of motion
we  can always construct three linear combinations of the form
\begin{equation}\label{s19}
\displaystyle{\mathcal{X}}'_{\alpha\alpha}=\int {d{\bf q}}\Big[\as{+}{\alpha,\zeta}({\bf q}){\sf a}^{-}_{\alpha,\zeta}({\bf q})+\as{-}{\alpha,\zeta}({\bf q}){\sf a}^{+}_{\alpha,\zeta}({\bf q})\Big],\quad \alpha=1,2,3.
\end{equation}
\noindent Therefore, the operators ${\sf
a}^{\pm}_{\alpha,\zeta}({\bf q}), \as{\pm}{\alpha,\zeta}({\bf q})$
lead to well-defined states in the Fock space.  The
difference between the number of particles and the number of
antiparticles of each type $\alpha$ with the same kinetic momentum
${\bf q}$ is an integral of motion.

Consider now the basis \eqref{001xx}. We can write the components of the field functions as
\begin{equation}\label{s20}
\begin{array}{c}
\displaystyle   {\Psi}^{(\nu)}_{s}(x)= \frac{1}{(2\pi)^{3/2}}\sum\limits_{l=1}^{3}
\sum\limits_{\zeta=\pm 1}\int \frac{d{\bf p}}{\sqrt{2p^{0}_{l}}}\,n^{(l)}_{s}\\[7pt] \displaystyle \;\;\;\times\left[
  e^{-\ii (px)}u_{p,\zeta}^{(m_l)}{\sf a}^{-}_{l,\zeta}({\bf p})+
  e^{\ii (px)}u_{-p,\zeta}^{(m_l)}{\sf a}^{+}_{l,\zeta}({\bf p})\right],\\[7pt]
\displaystyle   \bar{{\Psi}}^{(\nu)}_{s}(x)= \frac{1}{(2\pi)^{3/2}}\sum\limits_{l=1}^{3}
\sum\limits_{\zeta=\pm 1}\int \frac{d{\bf p}}{\sqrt{2p^{0}_{l}}}\,\na{l}{s}\\[7pt] \displaystyle \;\;\;\;\times\left[
  e^{-\ii (px)}\bar{u}_{-p,\zeta}^{(m_l)}\as{-}{l,\zeta}({\bf p})+
  e^{\ii (px)}\bar{u}_{p,\zeta}^{(m_l)}\as{+}{l,\zeta}({\bf p})\right].
\end{array}
\end{equation}
\noindent The operators  ${\sf a}^{\pm}_{l,\zeta}({\bf p}),
\as{\pm}{l,\zeta}({\bf p})$ arise as a result of the scaling
transformation
\begin{equation}\label{s21}
{\sf a}^{\pm}_{l,\zeta}({\bf p})=(\mu_{l})^{-3/2}{\sf a}^{\pm}_{l,\zeta}({\bf q}_{\,l}),\;\;\as{\pm}{l,\zeta}({\bf p})=(\mu_{l})^{-3/2}\as{\pm}{l,\zeta}
({\bf q}_{\,l}),\;\;{\bf q}_{\,l}={\bf p}/\mu_{l}.
\end{equation}
\noindent Therefore, using Eq. \eqref{s16} and  the invariance
condition with respect to the change of
particles to antiparticles for $ {\cal{P}}_{\beta} $, we get that these operators as well as
their linear combina\-tions
\begin{equation}\label{s22}
{\sf a}^{\pm}_{\alpha,\zeta}({\bf p})=\sum\limits_{l=1}^{3}U_{\alpha l}{\sf a}^{\pm}_{l,\zeta}({\bf p}),\;\; \as{\pm}{\alpha,\zeta}({\bf p})=\sum\limits_{l=1}^{3}U_{\alpha l}^{*}\as{\pm}{l,\zeta}({\bf p})
\end{equation}
\noindent satisfy the canonical commutation relations \eqref{s17}.

A similar reasoning shows that the states, which are described  by the
operators ${\sf a}^{\pm}_{l,\zeta}({\bf p})$, $
\as{\pm}{l,\zeta}({\bf p})$ and ${\sf a}^{\pm}_{\alpha,\zeta}({\bf
p}), \as{\pm}{\alpha,\zeta}({\bf p})$ are well-defined in the Fock
space, too. However, there is an important difference. The integral of
motion $ {\cal P}_{\beta}$ is non-diagonal  in terms of  the operators
${\sf a}^{\pm}_{\alpha,\zeta}({\bf p}),\as{\pm}
{\alpha,\zeta}({\bf p})$. This situation is quite
expected. The integral of motion $ {\cal P}_{\beta} $ is not the
canonical momentum  of the field,   but
the ``kinetic momentum''. Since the superpositions of the mass states are non-stationary
states, the non-diagonal form of $ {\cal P}_{\beta} $ reflects the
fact of a possible energy transfer from one state to another.

If we  admit that the matrices ${\mathds{V}}^{(i)} $ are
equal, we come, at least in the framework of the perturbation
theory, to a model with three independent  generations of
fermions. If the matrices $ {\mathds{V}}^{(i)} $ are distinct, the
terms of the Lagrangian that describe the interaction via the
charged currents immediately cause  a phenomenon, which is known
as  the mixing of generations. The matrix of the mixing
coefficients for  quarks is an analog  of the
Cabibbo--Kobayashi--Maskawa (CKM) matrix
\begin{equation}\label{l4}
{\mathds{U}}^{\mathrm{CKM}}={\mathds{V}}^{(u)\dag}
{\mathds{V}}^{(d)},
\end{equation}
\noindent and for  leptons it is  an analog  of the
Pontecorvo--Maki--Nakagawa--Sakata (PMNS) matrix
\begin{equation}\label{l5}
{\mathds{U}}^{\mathrm{PMNS}}={\mathds{V}}^{(e)\dag}{\mathds{V}}^{(\nu)}.
\end{equation}
\noindent The elements of these matrices can be expressed by the
scalar products of the basis vectors. For example,
\begin{equation}\label{18}
\sum\limits_{s=1}^{3}n^{(l)}_{s}{\ap{\alpha}{s}}
= U_{\alpha l}^{\mathrm{PMNS}}.
\end{equation}
\noindent  We shall use the
notation $U_{\alpha l }^{\mathrm{PMNS}}\equiv P_{\alpha l }$ for the
PMNS matrix elements to omit the  lengthy indices.

In the presence of the mixing, the transition probabilities
for the superpositions of the mass states cannot be reduced
to the sum of the transition probabilities for the mass states. It
should be emphasized that the total transition probabilities for
these states can be different for the same unitary matrix $
{\mathds{U}} $, but various complete sets of the operators
defining the  mass state. The reason is that the operators
\eqref{000020xxx} for such states can be different. Moreover,
the  measurement results can be different for different
space-time localization points of the detector.
In  experiment, as it is well known, the generation mixing occurs
for quarks, while the transfer of energy from one neutrino state to another is seen as the oscillation phenomenon.

\section{Neutrino oscillations}\label{M.3}

Now we try to apply the developed formalism to the problem of
neutrino oscillations. Eq. \eqref{6} is obviously translation
invariant. Therefore, shifting  the argument $ x^{\mu} $
in its solution  by a constant 4-vector $ z^{\mu} $, we still obtain  a solution.  For the solutions describing
the mass states, this transformation leads to a trivial phase
multiplier. However, the form of the solutions that are
determined by  formula \eqref{016x} will change. That is, an
argument shift generates a unitary transformation in the solution
space.

Let us examine this issue in detail. For clarity, we will
describe the neutrino with the help of  the density matrices of
pure states. By definition,  the density matrix
of a pure state is $\varrho(x,y)=\Psi(x){\bar\Psi(y)}$.
We assume that the neutrino source  and the detector  are at a
distance $ L$. In the area where a neutrino is produced, it is
described by the density matrix
$\varrho^{(\nu)}(x,y;q,\zeta,\alpha)$, and in the detection
area it is described by the density matrix
$\varrho^{(\nu)}(x-z,y-z;q,\zeta,\alpha)$.  A mass state can
be described, for example, by a density matrix
\begin{equation}\label{16}
\displaystyle \varrho^{(\nu)}(x,y;q,\zeta,\mu_{l}) = \frac{1}{{4{q}^{0}}}\, e^{-\ii(q(x-y))\mu_{l}}\mu_{l}^{3}\left(n^{(l)}
\otimes\na {l}{\phantom{j}}\right)(
\gamma_{\mu}{q}^{\mu} + m)(1 - \zeta\gamma ^5\gamma_{\mu} {S}_0^{\mu}(q)),
\end{equation}
\noindent which is based on the solutions \eqref{001x}, or by a density matrix
\begin{equation}\label{0016}
 \displaystyle\varrho^{(\nu)}(x,y;p,\zeta,m_{l}) = \frac{1}{4{p}^{\,0}_{l}}\, e^{-\ii(p_{l}(x-y))}\left(n^{(l)}\otimes\na {l}{\phantom{j}}\right)
(\gamma_{\mu}{p}^{\mu}_{l} + m_{l})(1 - \zeta
\gamma ^5\gamma_{\mu} {S}_0^{\mu}(p_{l})),
\end{equation}
\noindent which is based on the solutions  \eqref{001xx}. In both cases
there is no dependence on  $ L $.

The situation changes in the case of superpositions of the mass
states. If we take a state described by a superposition of the
wave functions \eqref{001x}, then the group velocities of all
the mass states of the neutrino will be the same: ${\bf v}
={\bf q}/q^{0}$. So we can suppose $z^{\mu}=q^{\mu}L/|{\bf q}|$.
Therefore, the density matrix describing the neutrino state at the
distance $ L $ from the  source can  be written as follows:
\begin{equation}\label{22}
\begin{array}{c}
   \displaystyle \varrho^{(\nu)}(x,y;q,\zeta,\mu_{l},\mu_{k};
   \alpha,L) = \frac{1}{4q^{0}}
    \Bigg[\sum\limits_{k,l=1}^{3}e^{-\ii(qx)
\mu_{l}+\ii(qy)\mu_{k}+2\pi \ii
L/L^{(lk)}}\\ \displaystyle \times
\left(\mu_{l}\mu_{k}\right)^{3/2}\left(n^{(l)}
\otimes\na {k}{\phantom{j}}\right) U_{\alpha
k}U_{\alpha l}^{*}\Bigg](
\gamma_{\mu}{q}^{\mu} + m)(1 - \zeta\gamma ^5\gamma_{\mu} {S}_0^{\mu}(q)).
\end{array}
\end{equation}
\noindent Here we introduced the notation
\begin{equation}\label{21}
L^{(lk)}= \frac{2\pi |{\bf q}|}{m(m_{l}-m_{k})}=\frac{2\pi \beta }{(m_{l}-m_{k})\sqrt{1-\beta^{2}}},\quad \beta =|{\bf v}|.
\end{equation}
\noindent Recall that in  formula \eqref{22}, as well as in
 formulas \eqref{16}, \eqref{0016}, the 4-vector $ q^{\mu}
$ satisfies the condition $q^{2}=m^{2}\!$, 4-vector $
{S}_0^{\mu}(q)$ determines the direction of  the polarization
of the particles, $ \zeta = \pm 1 $ is the  sign of the
spin projection on this direction. If we consider the density
matrix \eqref{22}, averaged over the parameters $\alpha $ and
$\zeta $, we obtain
\begin{equation}\label{23}
 \begin{array}{l}  \displaystyle  \varrho^{(\nu)}(x,y;q,\mu_{l}) = \displaystyle \!\frac{1}{12q^{0}}
    \!\left[\sum\limits_{l=1}^{3}e^{-\ii(q(x-y))
\mu_{l}}\mu_{l}^{3}\left(n^{(l)}\otimes\na {l}{\phantom{j}}\right)
\right]\!\!(\gamma_{\mu}{q}^{\mu} + m).
\end{array}
\end{equation}
\noindent It is clear that  the dependence of the density
matrix on the distance from the source appears only when the
source is coherent. This requires the dimensions of the source to
be small compared with  the  parameters $ {L}^{(lk)}$.

If we take a state described by a superposition of  wave
functions \eqref{001xx}, then the group velocities of the mass
states ${\bf v}_{l} ={\bf p}/p^{0}_{l}$ are different. However,
if $m_{l,k}^{2}/|{\bf p}|^{2} \ll 1 $, this
difference is extremely small, and we can set
$$z^{0}\approx L(p^{0}_{l}+p^{0}_{k})/(2|{\bf p}|)\approx Lp^{0}_{k}/|{\bf p}|\approx Lp^{0}_{l}/|{\bf p}|, $$
So
$$ z^{0}(p^{0}_{l}-p^{0}_{k}) =2\pi\big(L/ \tilde{L}^{(lk)}\big)\big(1+{\cal O}(m_{l,k}^{2}/|{\bf p}|^{2})\big),$$
where
\begin{equation}\label{k21}
\tilde{L}^{(lk)}= \frac{4\pi |{\bf p}|}{(m_{l}^{2}-m_{k}^{2})}\,.
\end{equation}
\noindent The density matrix describing the superposition of the
 neutrino mass states at  the  distance $ L $ from the source,
can
be calculated if one uses the  explicit form of the spinors $u_{p,\zeta}^{(m_{l})}$ (see \cite{LL}).
With the notation \eqref{k21}, we have
\begin{equation}\label{022}
\begin{array}{l}
  \displaystyle
 \varrho^{(\nu)}(x,y;p,\zeta,m_{l},m_{k};\alpha,L) \\ [8pt] 
   \displaystyle =\frac{1}{8}
    \sum\limits_{k,l=1}^{3}\frac{1}
    {\sqrt{p^{0}_{l}p^{0}_{k}}}\,e^{-\ii(p_{l}x)+
    \ii(p_{k}y)
    +2\pi\ii {L}/\tilde{L}^{(lk)}}
    \left(n^{(l)}\otimes\na {k}{\phantom{j}}\right) U_{\alpha
k}U_{\alpha l}^{*}
\\[12pt]\times\displaystyle (\gamma_{\mu}{p}^{\mu}_{l} + m_{l})(1 - \zeta\gamma ^5\gamma_{\mu} {S}_0^{\mu}(p_{l}) )\!\left[\sqrt{\frac{p_{k}^{0}+m_{k}}{p_{l}^{0}+m_{l}}}
\left(1+\gamma^{0}\right)
+\sqrt{\frac{p_{k}^{0}-m_{k}}
{p_{l}^{0}-m_{l}}}\left(1-\gamma^{0}\right)\right]
\\[12pt]
\equiv\displaystyle\frac{1}{8}
    \sum\limits_{k,l=1}^{3}\frac{1}
    {\sqrt{p^{0}_{l}p^{0}_{k}}}\,e^{-\ii(p_{l}x)+
    \ii(p_{k}y)
    +2\pi\ii {L}/\tilde{L}^{(kl)}}
    \left(n^{(l)}\otimes\na {k}{\phantom{j}}\right) U_{\alpha
k}U_{\alpha l}^{*}
\\[16pt]\times\displaystyle \left[\sqrt{\frac{p_{l}^{0}+m_{l}}{p_{k}^{0}+m_{k}}}
\left(1+\gamma^{0}\right)
+\sqrt{\frac{p_{l}^{0}-m_{l}}
{p_{k}^{0}-m_{k}}}\left(1-\gamma^{0}\right)\right]\!
(\gamma_{\mu}{p}^{\mu}_{k} + m_{k})(1 - \zeta\gamma ^5\gamma_{\mu} {S}_0^{\mu}(p_{k}) ).
\end{array}
\end{equation}
\noindent If we consider the density matrix \eqref{022}, averaged
over the parameters $\alpha $ and $\zeta $, we obtain
\begin{equation}\label{023}
\varrho^{(\nu)}(x,y;p,m_{l})=
    \displaystyle\sum\limits_{l=1}^{3}\frac{1}
    {12p^{0}_{l}}\,e^{-\ii(p_{l}(x-y))}
    \left(n^{(l)}\otimes\na {l}{\phantom{j}}\right)(\gamma_{\mu}{p}^{\mu}_{l} + m_{l}).
\end{equation}

Using \eqref{22} and \eqref{022} we can get the probability of
the neutrino transition from one state to another. These states
differ from each other, so for the state \eqref{22} the
decoherence effect is absent and the state \eqref{022} spreads due
to different group velocities of its components. Due to the fact
that the interaction conserves the energy, but not the group
velocity, as will be shown later, the approximate equality
(${\mathcal{E}}_{\nu}$ is the average value of the neutrino
energy)
\begin{equation}\label{024x}
\tilde{L}^{(lk)} \approx {L}^{(lk)}\approx {L}^{(lk)}_{osc}=\frac{4\pi {\mathcal{E}}_{\nu}}{(m_{l}^{2}-m_{k}^{2})}
\end{equation}
\noindent holds, and the condition of applicability  of  both
formulas \eqref{22} and \eqref{022} is defined by  the relation
 \begin{equation}\label{kkk21}
\frac{m^{2}_{l,k} L}{|{\bf p}|^{2}{L}^{(lk)}_{osc}}\ll 1.
\end{equation}

In accordance with  the fundamental principles
of quantum mechanics the probability of  observing the state
$\alpha$ at a distance $L$ from the production point  of the
state $\beta$ is
\begin{equation}\label{dens}
{\mathcal{P}}_{\nu_{(\beta)}\rightarrow\nu_{(\alpha)}}=
{\mathrm{Sp}}(\varrho(x,y;\alpha,L)\varrho^{\dag}(x,y;\beta)).
\end{equation}
\noindent Therefore, in both cases
\begin{equation}\label{36x}
{\mathcal{P}}_{\nu_{(\beta)}\rightarrow\nu_{(\alpha)}}=
\sum\limits_{k,l=1}^{3}U_{\beta
l}U_{\beta k}^{*}U_{\alpha k}U_{\alpha l}^{*}e^{2\pi \ii
L/L_{osc}^{(lk)}}.
\end{equation}
\noindent The formula itself is quite trivial. Mathematically it gives the
sum of the squared absolute values of the
projections of the new basis unit vectors in the
three-dimensional vector space over the field of complex numbers
onto the original basis vectors. In our case, the unitary
transformation of the  basis is generated by the phase
factors arising as a result of space-time translations. Regardless
of the initial basis, i.e. the parameters $ U_{\beta l} $, the
result is the same. Therefore,  formula \eqref{36x} describes the
evolution of an arbitrary superposition of the mass states. In
particular, we can consider the states that are
conventionally called the flavor states. Their
wave functions are defined by
the PMNS matrix $U_{\alpha l }^{\mathrm{PMNS}}\equiv P_{\alpha l }$.

In the phenomenological theory of neutrino oscillations
only the transitions between the flavor states are considered.
That is, it is postulated that the neutrino is produced  in a
flavor state. This postulate is quite natural, since the PMNS
matrix is the only non-trivial unitary $ 3\times 3 $ matrix that
can be built using the elements of the representation space
for leptons. However, as we have seen, the flavor states are not
uniquely defined. Their properties are associated with the type of
the mass states, from which they are composed. The only way to
answer  the question, in what state  the neutrinos are produced,
is to calculate the corresponding probabilities of the processes.

\section{Pion decay}\label{M.4}

In the processes involving neutrinos  only  the charged
leptons (the electron, the muon and the tau-lepton) are
detected directly. As the charged
leptons  have fixed masses $ M^{(\beta)} $,  their description should be done using the density
matrix of the mass states $\varrho^{(e)}(x,y;
p,\zeta,\eta_{\beta})$ or $
\varrho^{(e)}(x,y;p,\zeta,M_{\beta})$. These density matrices
are quite similar to the density matrices of neutrinos that
are given by \eqref{16}, \eqref{0016}.
The result of the total probability calculation does not depend
on the type of the density matrix. It is quite
obvious, since any mass state is defined by an eigenfunction of
the operator which is determined by the mass matrix.

In the neutral currents processes the oscillations
cannot be observed, so we can sum directly  over all the discrete
quantum numbers  of the neutrino final states using
either the density matrix \eqref{23} or \eqref{023} (multiplied by
6) in the calculations. Thus, the probability does not depend on
$ L $.

Therefore, we will examine how neutrino oscillations affect the
probability  of the processes which occur via the charged
currents only. As an example, consider the pion decay
\begin{equation*}\label{25}
    {{\pi ^{+}\Rightarrow l^{+}_{\beta}+\nu}},\qquad l^{+}_{\beta}= \mu^{+}, e^{+}.
\end{equation*}

Let the 4-momentum of the pion be $ k^{\mu}, \, k^{2} =
m_{\pi}^{2} $, \, and 4-momentum of  the lepton $ l^{+}_{\beta}
$ be $ p^{\mu}, \, p^{2} = M_{\beta}^{2} $. We assume that the
distance from the source of neutrinos is  $ L $, and the linear
size of the source is  $ L_{0} $. For clarity, we assume that
the pion is at rest: $k^{0}=m_{\pi},\,{\bf k}=0$.

First, consider the  probability of the process that produces a
charged lepton with mass $ M_{\beta} $ and a  neutrino with mass $
m_{l} $. The  probability of this process  in the Fermi
approximation is given by the formula
\begin{equation}\label{x16}
\begin{array}{l}
\displaystyle W_{\beta l}=
\frac{G^{2}_{\mathrm{F}}f_{\pi}^{2}}{4(2\pi)^{6}
k^{0}}\int d^{4}x
d^{4}y\int d{\bf q}d{\bf p}\\[6pt]\displaystyle \times{\mathrm
{Sp}}\Big\{\varrho^{(e)}(x,y;-p,\zeta,\beta)
\gamma^{\mu}(1+\gamma^{5})
{\varrho}^{(\nu)}(y,x;q,\zeta,{l})
\gamma^{\nu}(1+\gamma^{5}) k_{\mu}k_{\nu}e^{-\ii(k(x-y))}\Big\}.
\end{array}
\end{equation}
\noindent  The density matrix of the neutrino
${\varrho}^{(\nu)}(y,x;q,\zeta,{l})$ can be taken either in
the form  \eqref{16} or \eqref{0016}. The density matrix of the charged antilepton $\varrho^{(e)}(x,y;-p,\zeta,\beta)$ can
be  taken either in the form  analogous to \eqref{g16}
\begin{equation}\label{17}
\displaystyle \varrho^{(e)}(x,y;-p,\zeta,\eta_\beta) = \frac{1}{4p^{0}}\, e^{\ii(p(x-y)){{\eta}}_{\beta}}{{\eta}}_{\beta}^{3}
\left(a^{(\beta)}
\otimes\ap{\beta}{\phantom{j}}\right)
(\gamma_{\mu}{p}^{\mu} - M)(1 -
\zeta\gamma ^5\gamma_{\mu}{S}_0^{\mu}(p)),
\end{equation}
\noindent or in the form analogous to  \eqref{g0016}
\begin{equation}\label{0017}
\displaystyle \varrho^{(e)}(x,y;-p,\zeta,M_\beta) = \frac{1}{4{p}^{\,0}_{\beta}}\, e^{\ii(p_{\beta}(x-y))}\left(a^{(\beta)}
\otimes\ap{\beta}{\phantom{j}}\right)
(\gamma_{\mu}{p}_{\beta}^{\mu} - M_{\beta})(1 - \zeta\gamma ^5 \gamma_{\mu}{S}_0^{\mu}(p_{\beta})).
\end{equation}

Substituting these expressions into \eqref{x16} and making elementary calculations, we obtain
\begin{equation}\label{x21.0}
\begin{array}{l}
\displaystyle W_{\beta l}=\frac{G^{2}_{\mathrm{F}}f_{\pi}^{2}}
{8\pi{m_{\pi}^{3}}}
P_{\beta l}P_{\beta l}^{*}\sqrt{\left(m_{\pi}^{2}-M_{\beta}^{2}+
m_{l}^{2}\right)^{2}
-4m_{\pi}^{2}m_{l}^{2}}\\[6pt] \displaystyle  \times\Big[M_{\beta}^{2}({m_{\pi}^{2}}-
{M_{\beta}^{2}}+m_{l}^{2})+
m_{l}^{2}({m_{\pi}^{2}}+{M_{\beta}^{2}}-
m_{l}^{2})\Big].
\end{array}
\end{equation}
\noindent This expression shows that the total probability of the pion decay,
\begin{equation}\label{x21.0z}
\displaystyle W_{\beta}=\sum\limits_{l=1}^{3}W_{l\beta}=
\frac{G^{2}_{\mathrm{F}}f_{\pi}^{2}}{8\pi{m_{\pi}^{3}}}
M_{\beta}^{2}
{\left({m_{\pi}^{2}}-{M_{\beta}^{2}}\right)^{2}}
\bigg(1+
\sum\limits_{l=1}^{3}{\mathcal{O}}
(P_{\beta l}P_{\beta l}^{*}m_{l}^{2}/{\mathcal{E}}_{\nu}^{2})\bigg),
\end{equation}
\noindent weakly depends on the mixing parameters $ P_{\beta l} $ (here ${\mathcal{E}}_{\nu}\approx ({m_{\pi}^{2}}-{M_{\beta}^{2}})/2m_{\pi}$).

Now we calculate the probability of detecting a superposition  of
the neutrino mass states  produced in the pion  decay with
the emission of a charged lepton with  mass $ M_{\beta} $ at a
distance $ L $ from the source. This probability is equal to
\begin{equation}\label{26}
\begin{array}{l}
\displaystyle W_{\beta\alpha}^{L}=
\frac{G^{2}_{\mathrm{F}}f_{\pi}^{2}}{4(2\pi)^{6}k^{0}}
\int d^{4}x
d^{4}y\int d{\bf q}d{\bf p}\\[6pt]\displaystyle
\times{\mathrm{Sp}}\Big\{\varrho^{(e)}(x,y;-p,
\zeta,\beta)
\gamma^{\mu}(1+\gamma^{5})
{\varrho}^{(\nu)}(y,x;q,\zeta,\alpha;L)
\gamma^{\nu}(1+\gamma^{5}) k_{\mu}k_{\nu}e^{-\ii(k(x-y))}\Big\}.
\end{array}
\end{equation}

First,  perform the calculations taking  the density
matrix ${\varrho}^{(\nu)}
(y,x;q,\zeta,\alpha; L) $ given by  formula \eqref{22}. That is, we assume that the group
velocities of the mass states are the same. We change the
integration variables
\begin{equation}\label{28}
 z_{-}^{\mu}= x^{\mu}- y^{\mu},\quad z_{+}^{\mu} = (x^{\mu}+ y^{\mu})/2,
\end{equation}
\noindent and limit the  range of integration over the variable $
z_{+}^{\mu}$ to the size of the area in which the reaction takes
place \cite{Schwinger}.  Then, after dividing by the volume of the
reaction area,  we obtain (see 
\ref{A.2})
\begin{equation}\label{29}
\begin{array}{c}
\displaystyle
W_{\beta\alpha}^{L}=\frac{G^{2}_{\mathrm{F}}f_{\pi}^{2}}
{8\pi{m_{\pi}^{3}}}
M_{\beta}^{2}
{\left({m_{\pi}^{2}}-{M_{\beta}^{2}}\right)^{2}}
\\[4pt] \displaystyle\times
\Bigg[\sum\limits_{k,l=1}^{3}\frac{ P_{\beta l}P_{\beta
k}^{*}U_{\alpha k}U_{\alpha l}^{*}}{\Delta^{3}_{lk}}\,\frac{\sin(\pi
L_{0}/L_{osc}
    ^{(lk)})}{\pi L_{0}/L_{osc}
    ^{(lk)}}e^{2\pi \ii
L/L_{osc}^{(lk)}}\Bigg] 
\Big(1+{\mathcal{O}}
(m_{l,k}^{2}/{\mathcal{E}}_{\nu}^{2})\Big).
\end{array}
\end{equation}
\noindent Here
\begin{equation}\label{32}
\Delta_{lk}= (m_{l}+m_{k})/(2\sqrt{m_{l}m_{k}})\geqslant 1
\end{equation}
\noindent is the ratio of the arithmetic mean to the geometric mean of
the corresponding neutrino masses. The origin of the factor
\begin{equation}\label{31}
R=\frac{\sin(\pi L_{0}/L_{osc}^{(lk)})}{\pi L_{0}/L_{osc}^{(lk)}}
\end{equation}
\noindent is the incomplete coherence of the source that is due to its
finite size $ L_{0}\ll L $. Obviously, (see \eqref{x40}), if $l=k$  we have
$R\equiv 1$ for any $L_{0}$.
After summation over the parameter
$ \alpha $ in  formula \eqref{29}, there appears an expression for
the total probability of the pion decay \eqref{x21.0z}.

Consider the probability of detecting the neutrino flavor
states at a close range $ L\ll|L_{osc}^{(lk)}| $ if the source is quite compact
$ L_{0}\ll L \ll|L_{osc}^{(lk)}| $.
Then, replacing
 $U_{\alpha l}$  by $P_{\alpha l}$ we obtain
\begin{equation}\label{34}
\displaystyle
W_{\beta\alpha}=\frac{G^{2}_{\mathrm{F}}f_{\pi}^{2}}
{8\pi{m_{\pi}^{3}}}
M_{\beta}^{2}
{\left({m_{\pi}^{2}}-{M_{\beta}^{2}}\right)^{2}}
\Bigg[\sum\limits_{k,l=1}^{3}\frac{P_{\beta l}P_{\beta
k}^{*}P_{\alpha k}P_{\alpha l}^{*}}{{\Delta^{3}_{kl}}}\Bigg]\Big(1+{\mathcal{O}}
(m_{l,k}^{2}/{\mathcal{E}}_{\nu}^{2})\Big).
\end{equation}
\noindent This formula shows that not only the probability of
detecting  the main neutrino flavor $ \beta $ is nonzero, but the
probability of detecting the other flavors is nonzero as well.
Thus, the state of the neutrino, that was produced in the decay
is a superposition of the flavor states described by the
density matrices \eqref{22}. It should be emphasized that we do
not talk about a very small contribution  $\sim
m_{l,k}^{2}/{\mathcal{E}}_{\nu}^{2}$.

Now take the density  matrix \eqref{022} for ${\varrho}^{(\nu)}\!(y,x;q,\zeta,\alpha;L)\!$,
that is, we assume
that the mass states included in the superposition have the same
canonical momenta. Using the same assumptions as in the previous
case, in particular, the change of variables \eqref{28} we obtain
\begin{equation}\label{37}
\begin{array}{c}
\displaystyle
W_{\beta\alpha}^{L}=\frac{G^{2}_{\mathrm{F}}f_{\pi}^{2}}
{8\pi{m_{\pi}^{3}}}
M_{\beta}^{2}
{\left({m_{\pi}^{2}}-{M_{\beta}^{2}}\right)^{2}} \\ \times\displaystyle
\Bigg[\sum\limits_{k,l=1}^{3}{ P_{\beta l}P_{\beta
k}^{*}U_{\alpha k}U_{\alpha l}^{*}}\,\frac{\sin(\pi
L_{0}/L_{osc}
    ^{(lk)})}{\pi L_{0}/L_{osc}
    ^{(lk)}}\,e^{2\pi \ii
L/L_{osc}^{(lk)}}\Bigg] 
\Big(1+{\mathcal{O}}
(m_{l,k}^{2}/{\mathcal{E}}_{\nu}^{2})\Big).
\end{array}
\end{equation}
\noindent When detecting the flavor states $( U_{\alpha l} =
P_{\alpha l}) $ at small distances from the compact source
\begin{equation}\label{35}
\displaystyle
W_{\beta\alpha}=
\frac{G^{2}_{\mathrm{F}}f_{\pi}^{2}}
{8\pi{m_{\pi}^{3}}}M_{\beta}^{2}
{\left({m_{\pi}^{2}}-{M_{\beta}^{2}}\right)^{2}}
\Big(\delta_{\alpha\beta}+{\mathcal{O}}
(m_{l,k}^{2}/{\mathcal{E}}_{\nu}^{2})\Big).
\end{equation}

Therefore, in the decay process  the flavor state $\beta$, which  is described with a good
accuracy by the density matrix \eqref{022} is
produced. This state can be
called  a flavor state in the conventional sense. Naturally, its
evolution is described by the standard formula
\begin{equation}\label{36}
{\mathcal{P}}_{\nu_{(\beta)}\rightarrow\nu_{(\alpha)}}=
\sum\limits_{k,l=1}^{3}P_{\beta
l}P_{\beta k}^{*}P_{\alpha k}P_{\alpha l}^{*}e^{2\pi \ii
L/L_{osc}^{(lk)}},
\end{equation}
\noindent which is analogous to  formula \eqref{36x}.

The question, what exact state is produced  in the weak
decays of particles, is still open. In our model, this state is
pure in the quantum-mechanical sense, so that when building this
state we need to vary, in general, all the quantum numbers of the
final state of the neutrino. If this state could be expressed as
a finite set of plane waves, then we would have to choose two
appropriate  matrices: a Hermitian matrix $
{{\mathds{L}}}^{(\nu)}$ (see. \eqref{016}) and a unitary matrix $
{{\mathds{U}}}^{(\nu)}$ (see. \eqref{016x}). It is not improbable that the
required state  can  be a wave packet, although not an abstract Gaussian wave
packet, but a wave packet with characteristics
explicitly depending on the type of the process in which it
occurs. However, this is not essential, if
the condition \eqref{kkk21}  holds.

It should be recalled that all particles are produces in wave packet states. This circumstance helps to explain the phenomenon of the oscillations. However, there is no need to introduce wave packets explicitly to describe this phenomenon. The restriction of the space-time domain of integration with  respect to the variable
$z_{+}^{\mu}$ (see Eq. \eqref{28}) makes it possible to  take into account the nonmonochromaticity of the  produced  particles. The oscillation lengths of charged leptons are very small due to their relatively large masses. The factor $R$ (see Eq. \eqref{31}) for such particles is negligible and the oscillations are non observable. Therefore, the technique used is essential only for the light particles, i.e. neutrinos.

When detecting the neutrino states  from  a very large
$(L_{0}\gg |L^{(lk)}_{osc}|)$ source we have
\begin{equation}\label{38}
\displaystyle W_{\beta\alpha}=\frac{G^{2}_{\mathrm{F}}f_{\pi}^{2}}
{8\pi{m_{\pi}^{3}}}
M_{\beta}^{2}
{\left({m_{\pi}^{2}}-{M_{\beta}^{2}}\right)^{2}}
\sum\limits_{l=1}^{3}{ P_{\beta l}P_{\beta
l}^{*}U_{\alpha l}U_{\alpha l}^{*}}\Big(1+{\mathcal{O}}(|L^{(lk)}_{osc}|/L_{0})+{\mathcal{O}}
(m_{l,k}^{2}/{\mathcal{E}}_{\nu}^{2})
\Big).
\end{equation}
\noindent In this case oscillations are absent. If $U_{\alpha l}=\delta_{\alpha l}$
we get  the  probability of the process that produces a
charged lepton with mass $ M_{\beta} $ and a neutrino with mass $m_{\alpha}$. And
if $U_{\alpha l}=P_{\alpha l}$ we get
the  probability of the process that produces a
charged lepton with mass $ M_{\beta} $ and a neutrino flavor state ${\alpha}$.

Similarly, we can consider the decay
\begin{equation*}\label{250}
    {\pi ^{-}\Rightarrow l^{-}_{\beta}+\bar{\nu}},\quad l^{-}_{\beta}= \mu^{-}, e^{-},
\end{equation*}
\noindent  using the density matrix
for antineutrinos \eqref{g16}, \eqref{g0016}, \eqref{g22},
\eqref{g022}, the corresponding formulas can be obtained by
replacing the sign of the oscillation length in the expression for
the probability, that is, by replacing  $ L^{(lk)}_{osc}$ with
$ L^{(kl)}_{osc} $. As is well known, this difference in the
formulas for the probabilities of the processes involving
particles and antiparticles can indicate  CP violation in the
theory.

Thus,  our  approach  adequately  describes  the  pheno\-me\-non of neutrino
oscillations, which  is, in fact, the pheno\-me\-non of the
oscillation of the probabilities  of the processes involving
neutrinos.
The flavor state, constructed as a superposition of the mass
states with the same momentum, is distinguished by the fact that
it is very close to the state which is formed in the decay
process.
The possibility to use  such
flavor states in order to  describe
neutrinos when the distance from the source is much larger than the
oscillation length, directly follows from the smallness of the
neutrino masses compared to their energies. Therefore, for ultra-relativistic  neutrinos $
(m_{l,k}^{2}/{\mathcal{E}}_{\nu}^{2}\ll 1)$, which are really
observed, the phenomenological theory
of oscillations based on the ideas of Pontecorvo (see, e.g., \cite{bp}),
is a very good approximation  for  quantum field theory.

\section{Conclusions}\label{M.5}

In this paper we put forward a modification of the electroweak
interaction theory, in which the fermions with the same
electroweak quantum numbers are placed in  fermion multiplets and
are treated as different quantum states of a single particle. That
is, in describing  the  electroweak interactions it is possible to
use four fundamental fermions only. In this model, the mixing and
oscillations of the particles arise as a direct consequence of the
general principles of quantum field theory. This approach enables
one  to calculate the probabilities of the processes taking place
in  detectors at long distances from the  source.
Calculations of higher order processes including the computation
of the contributions due to radiative corrections can be performed
in the framework of perturbation theory using the regular diagram
technique.

The developed approach is used to study neutrino oscillations.
It is shown by the  example of the pion decay that the states
of the ultra-relativistic neutrino produced in the decay process can be described
by a superposition of states with different masses and identical
canonical momenta  with very high accuracy.

\section*{Acknowledgments}

Author is grateful to A.B. Arbuzov, A.V.
Bol\-sinov, A.V. Borisov, A.V. Chukhnova, A.D. Dolgov, D.V.~Gal'tsov,  A.V. Kartavtsev, E.M. Murchikova, A.A. Po\-lo\-sin, A.A. Slavnov, A.E. Shabad, I.P. Volobuev, and V.Ch. Zhu\-kovsky for numerous consultations and fruitful discussions.

\appendix


\section{Explicit form for the generators  of
$SU(3)$ transformations}\label{A.0}

A direct calculation yields  the explicit form of the generators
of $SU(3)$ transformations ${X}_{k}^{(\nu)}={\cal K}^{(\nu)}{X}_{k}{\cal K}^{(\nu)-1}$
\vspace{-5pt}
\begin{equation}\label{aaa2}
\begin{array}{l}
\displaystyle {X}_{1}^{(\nu)}=\left(n^{(1)}\otimes \na{2}{\phantom{j}}\right){\cal D}_{(1)}{\cal D}_{(2)}^{-1}+\left(n^{(2)}\otimes \na{1}{\phantom{j}}\right){\cal D}_{(2)}{\cal D}_{(1)}^{-1}, \\[4pt]
\displaystyle {X}_{2}^{(\nu)}=\ii\left(n^{(2)}\otimes \na{1}{\phantom{j}}\right){\cal D}_{(2)}{\cal D}_{(1)}^{-1}-\ii\left(n^{(1)}\otimes \na{2}{\phantom{j}}\right){\cal D}_{(1)}{\cal D}_{(2)}^{-1}, \\[4pt]
\displaystyle {X}_{3}^{(\nu)}=\left(n^{(1)}\otimes \na{1}{\phantom{j}}\right)-\left(n^{(2)}\otimes \na{2}{\phantom{j}}\right), \\[4pt]
\displaystyle {X}_{4}^{(\nu)}=\left(n^{(1)}\otimes \na{3}{\phantom{j}}\right){\cal D}_{(1)}{\cal D}_{(3)}^{-1}+\left(n^{(3)}\otimes \na{1}{\phantom{j}}\right){\cal D}_{(3)}{\cal D}_{(1)}^{-1}, \\[4pt]
\displaystyle {X}_{5}^{(\nu)}=\ii\left(n^{(3)}\otimes \na{1}{\phantom{j}}\right){\cal D}_{(3)}{\cal D}_{(1)}^{-1}-\ii\left(n^{(1)}\otimes \na{3}{\phantom{j}}\right){\cal D}_{(1)}{\cal D}_{(3)}^{-1}, \\[4pt]
\displaystyle {X}_{6}^{(\nu)}=\left(n^{(2)}\otimes \na{3}{\phantom{j}}\right){\cal D}_{(2)}{\cal D}_{(3)}^{-1}+\left(n^{(3)}\otimes \na{2}{\phantom{j}}\right){\cal D}_{(3)}{\cal D}_{(2)}^{-1}, \\[4pt]
\displaystyle {X}_{7}^{(\nu)}=\ii\left(n^{(3)}\otimes \na{2}{\phantom{j}}\right){\cal D}_{(3)}{\cal D}_{(2)}^{-1}-\ii\left(n^{(2)}\otimes \na{3}{\phantom{j}}\right){\cal D}_{(2)}{\cal D}_{(3)}^{-1}, \\[4pt]
\displaystyle {X}_{8}^{(\nu)}=\left(n^{(1)}\otimes \na{1}{\phantom{j}}\right)+\left(n^{(2)}\otimes \na{2}{\phantom{j}}\right)-2\left(n^{(3)}\otimes \na{3}{\phantom{j}}\right).
\end{array}
\end{equation}
Operators of finite transformations are given by
\begin{equation}\label{aaa4}
\begin{array}{l}
\displaystyle {U}_{1}^{(\nu)}(\alpha)=\left(n^{(3)}\otimes \na{3}{\phantom{j}}\right)+\left[ \left(n^{(1)}\otimes \na{1}{\phantom{j}}\right)+\left(n^{(2)}\otimes \na{2}{\phantom{j}}\right)\right]\cos\alpha-\ii{X}_{1}^{(\nu)}
\sin\alpha, \\[4pt]
\displaystyle {U}_{2}^{(\nu)}(\alpha)=\left(n^{(3)}\otimes \na{3}{\phantom{j}}\right)+\left[ \left(n^{(1)}\otimes \na{1}{\phantom{j}}\right)+\left(n^{(2)}\otimes \na{2}{\phantom{j}}\right)\right]\cos\alpha-\ii{X}_{2}^{(\nu)}
\sin\alpha, \\[4pt]
\displaystyle {U}_{3}^{(\nu)}(\alpha)=\left(n^{(3)}\otimes \na{3}{\phantom{j}}\right)+\left(n^{(1)}\otimes \na{1}{\phantom{j}}\right)e^{-\ii\alpha}+\left(n^{(2)}\otimes \na{2}{\phantom{j}}\right)e^{\ii\alpha}, \\[4pt]
\displaystyle {U}_{4}^{(\nu)}(\alpha)=\left(n^{(2)}\otimes \na{2}{\phantom{j}}\right)+\left[ \left(n^{(1)}\otimes \na{1}{\phantom{j}}\right)+\left(n^{(3)}\otimes \na{3}{\phantom{j}}\right)\right]\cos\alpha-\ii{X}_{4}^{(\nu)}
\sin\alpha, \\[4pt]
\displaystyle {U}_{5}^{(\nu)}(\alpha)=\left(n^{(2)}\otimes \na{2}{\phantom{j}}\right)+\left[ \left(n^{(1)}\otimes \na{1}{\phantom{j}}\right)+\left(n^{(3)}\otimes \na{3}{\phantom{j}}\right)\right]\cos\alpha-\ii{X}_{5}^{(\nu)}
\sin\alpha, \\[4pt]
\displaystyle {U}_{6}^{(\nu)}(\alpha)=\left(n^{(1)}\otimes \na{1}{\phantom{j}}\right)+\left[ \left(n^{(2)}\otimes \na{2}{\phantom{j}}\right)+\left(n^{(3)}\otimes \na{3}{\phantom{j}}\right)\right]\cos\alpha-\ii{X}_{6}^{(\nu)}
\sin\alpha, \\[4pt]
\displaystyle {U}_{7}^{(\nu)}(\alpha)=\left(n^{(1)}\otimes \na{1}{\phantom{j}}\right)+\left[ \left(n^{(2)}\otimes \na{2}{\phantom{j}}\right)+\left(n^{(3)}\otimes \na{3}{\phantom{j}}\right)\right]\cos\alpha-\ii{X}_{7}^{(\nu)}
\sin\alpha, \\[4pt]
\displaystyle {U}_{8}^{(\nu)}(\alpha)=\left(n^{(3)}\otimes \na{3}{\phantom{j}}\right)e^{2\ii\alpha}+\left[ \left(n^{(1)}\otimes \na{1}{\phantom{j}}\right)+\left(n^{(2)}\otimes \na{2}{\phantom{j}}\right)\right]e^{-\ii\alpha},
\end{array}
\end{equation}
\noindent where
\begin{equation}\label{aaa3}
\begin{array}{l}
\displaystyle \left(n^{(1)}\otimes \na{1}{\phantom{j}}\right)= \frac{1}{6}\big(2{\mathds{I}}+{X}_{8}^{(\nu)}+3{X}_{3}^{(\nu)}\big), \\[8pt]
\displaystyle \left(n^{(2)}\otimes \na{2}{\phantom{j}}\right)= \frac{1}{6}\big(2{\mathds{I}}+{X}_{8}^{(\nu)}-3{X}_{3}^{(\nu)}\big), \\[8pt]
\displaystyle \left(n^{(3)}\otimes \na{3}{\phantom{j}}\right)= \frac{1}{3}\big({\mathds{I}}-{X}_{8}^{(\nu)}\big).
\end{array}
\end{equation}

\noindent Integrals of motion
\begin{equation}\label{s7}
    {\mathcal{X}}_{3}=\sum\limits_{\zeta=\pm 1}\int {d{\bf q}}\Big[\as{+}{1,\zeta}({\bf q}){\sf a}^{-}_{1,\zeta}({\bf q})+\as{-}{1,\zeta}({\bf q}){\sf a}^{+}_{1,\zeta}({\bf q})-\as{+}{2,\zeta}({\bf q}){\sf a}^{-}_{2,\zeta}({\bf q})-\as{-}{2,\zeta}({\bf q}){\sf a}^{+}_{2,\zeta}({\bf q})\Big],
\end{equation}
\begin{equation}\label{s8}
\begin{array}{l}
  \displaystyle{\mathcal{X}}_{8}=\sum\limits_{\zeta=\pm 1}\int {d{\bf q}}\Big[\as{+}{1,\zeta}({\bf q}){\sf a}^{-}_{1,\zeta}({\bf q})+\as{-}{1,\zeta}({\bf q}){\sf a}^{+}_{1,\zeta}({\bf q})+\as{+}{2,\zeta}({\bf q}){\sf a}^{-}_{2,\zeta}({\bf q})+\as{-}{2,\zeta}({\bf q}){\sf a}^{+}_{2,\zeta}({\bf q}) \\[20pt] \displaystyle\phantom{,\sum\limits_{\zeta=\pm 1}\int {d{\bf q}}}-2\as{+}{3,\zeta}({\bf q}){\sf a}^{-}_{3,\zeta}({\bf q})-2\as{-}{3,\zeta}({\bf q}){\sf a}^{+}_{3,\zeta}({\bf q})\Big],
\end{array}
\end{equation}
\begin{equation}\label{s9}
    {\mathcal{X}}_{1}=\sum\limits_{\zeta=\pm 1}\int {d{\bf q}}\Big[\as{+}{1,\zeta}({\bf q}){\sf a}^{-}_{2,\zeta}({\bf q})+\as{-}{1,\zeta}({\bf q}){\sf a}^{+}_{2,\zeta}({\bf q})+\as{+}{2,\zeta}({\bf q}){\sf a}^{-}_{1,\zeta}({\bf q})+\as{-}{2,\zeta}({\bf q}){\sf a}^{+}_{1,\zeta}({\bf q})\Big],
\end{equation}
\begin{equation}\label{s10}
    {\mathcal{X}}_{2}=\ii\sum\limits_{\zeta=\pm 1}\int {d{\bf q}}\Big[\as{+}{1,\zeta}({\bf q}){\sf a}^{-}_{2,\zeta}({\bf q})+\as{-}{1,\zeta}({\bf q}){\sf a}^{+}_{2,\zeta}({\bf q})-\as{+}{2,\zeta}({\bf q}){\sf a}^{-}_{1,\zeta}({\bf q})-\as{-}{2,\zeta}({\bf q}){\sf a}^{+}_{1,\zeta}({\bf q})\Big],
\end{equation}
\begin{equation}\label{s11}
   {\mathcal{X}}_{4}=\sum\limits_{\zeta=\pm 1}\int {d{\bf q}}\Big[\as{+}{1,\zeta}({\bf q}){\sf a}^{-}_{3,\zeta}({\bf q})+\as{-}{1,\zeta}({\bf q}){\sf a}^{+}_{3,\zeta}({\bf q})+\as{+}{3,\zeta}({\bf q}){\sf a}^{-}_{1,\zeta}({\bf q})+\as{-}{3,\zeta}({\bf q}){\sf a}^{+}_{1,\zeta}({\bf q})\Big],
\end{equation}
\begin{equation}\label{s12}
    {\mathcal{X}}_{5}=\ii\sum\limits_{\zeta=\pm 1}\int {d{\bf q}}\Big[\as{+}{1,\zeta}({\bf q}){\sf a}^{-}_{3,\zeta}({\bf q})+\as{-}{1,\zeta}({\bf q}){\sf a}^{+}_{3,\zeta}({\bf q})-\as{+}{3,\zeta}({\bf q}){\sf a}^{-}_{1,\zeta}({\bf q})-\as{-}{3,\zeta}({\bf q}){\sf a}^{+}_{1,\zeta}({\bf q})\Big],
\end{equation}
\begin{equation}\label{s13}
   {\mathcal{X}}_{6}=\sum\limits_{\zeta=\pm 1}\int {d{\bf q}}\Big[\as{+}{2,\zeta}({\bf q}){\sf a}^{-}_{3,\zeta}({\bf q})+\as{-}{2,\zeta}({\bf q}){\sf a}^{+}_{3,\zeta}({\bf q})+\as{+}{3,\zeta}({\bf q}){\sf a}^{-}_{2,\zeta}({\bf q})+\as{-}{3,\zeta}({\bf q}){\sf a}^{+}_{2,\zeta}({\bf q})\Big],
\end{equation}
\begin{equation}\label{s14}
   {\mathcal{X}}_{7}=\ii\sum\limits_{\zeta=\pm 1}\int {d{\bf q}}\Big[\as{+}{2,\zeta}({\bf q}){\sf a}^{-}_{3,\zeta}({\bf q})+\as{-}{2,\zeta}({\bf q}){\sf a}^{+}_{3,\zeta}({\bf q})-\as{+}{3,\zeta}({\bf q}){\sf a}^{-}_{2,\zeta}({\bf q})-\as{-}{3,\zeta}({\bf q}){\sf a}^{+}_{2,\zeta}({\bf q})\Big].
\end{equation}

\section{Density matrix of antineutrino}\label{A.1}

The mass states of antineutrino are described either by the
density matrix
\begin{equation}\label{g16}
\displaystyle \varrho^{(\nu)}(x,y;-q,\zeta,\mu_{l}) = \frac{1}{{4{q}^{0}}}\, e^{-\ii(q(x-y))\mu_{l}}\mu_{l}^{3}\left(n^{(l)}\otimes\na {l}{\phantom{j}}\right)(
\gamma_{\mu}{q}^{\mu}- m)(1 - \zeta\gamma ^5\gamma_{\mu} {S}_0^{\mu}(q)),
\end{equation}
\noindent based on the solutions similar to \eqref {001x}, or the density matrix
\begin{equation}\label{g0016}
\displaystyle \varrho^{(\nu)}(x,y;-p,\zeta,m_{l}) = \frac{1}{4{p}^{\,0}_{l}}\, e^{\ii(p_{l}(x-y))}\left(n^{(l)}\otimes\na {l}{\phantom{j}}\right)
(\gamma_{\mu}{p}^{\mu}_{l} - m_{l})(1 - \zeta\gamma ^5\gamma_{\mu} {S}_0^{\mu}(p_{l})).
\end{equation}
\noindent based on the solutions  similar to \eqref{001xx}. In both cases, there is no dependence on $ L $.

If we consider a state, which is described by a superposition of wave
functions similar to \eqref{001x}, then the density matrix for the
state of an antineutrino at a distance $ L $ from the
source can be written as
\vspace{-4pt}
\begin{equation}\label{g22}
\begin{array}{l}
\displaystyle \varrho^{(\nu)}(x,y;-q,\zeta,\mu_{l},\mu_{k};
   \alpha,L)  = \displaystyle\frac{1}{4q^{0}}
    \Bigg[\sum\limits_{k,l=1}^{3}e^{\ii(qx)
\mu_{l}-\ii(qy)\mu_{k}+2\pi \ii
L/L^{(lk)}}\\
\displaystyle \phantom{dddddd}\times
\left(\mu_{l}\mu_{k}\right)^{3/2}\left(n^{(l)}
\otimes\na {k}{\phantom{j}}\right) U_{\alpha
k}U_{\alpha l}^{*}\Bigg](
\gamma_{\mu}{q}^{\mu} - m)(1 - \zeta\gamma ^5\gamma_{\mu} {S}_0^{\mu}(q)).
\end{array}
\end{equation}
If we consider a state  described by a superposition of wave
functions similar to \eqref{001xx}, then the density matrix for
the state of an antineutrino at a distance $ L $ from the
source can be written as
\begin{equation}\label{g022}
\begin{array}{l}
\displaystyle
  \varrho^{(\nu)}(x,y;-p,\zeta,m_{l},m_{k};\alpha,L)\\ [8pt] 
   \displaystyle =\frac{1}{8}
    \sum\limits_{k,l=1}^{3}\frac{1}
    {\sqrt{p^{0}_{l}p^{0}_{k}}}\,e^{\ii(p_{l}x)-
    \ii(p_{k}y)
    +2\pi\ii {L}/\tilde{L}^{(lk)}}
    \left(n^{(l)}\otimes\na {k}{\phantom{j}}\right) U_{\alpha
k}U_{\alpha l}^{*}
\\[12pt]\times\displaystyle (\gamma_{\mu}{p}^{\mu}_{l} - m_{l})(1 - \zeta\gamma ^5\gamma_{\mu} {S}_0^{\mu}(p_{l}) )\!\left[\sqrt{\frac{p_{k}^{0}-m_{k}}{p_{l}^{0}-m_{l}}}
\left(1+\gamma^{0}\right)
+\sqrt{\frac{p_{k}^{0}+m_{k}}
{p_{l}^{0}+m_{l}}}\left(1-\gamma^{0}\right)\right]
\\[12pt]
\equiv\displaystyle\frac{1}{8}
    \sum\limits_{k,l=1}^{3}\frac{1}
    {\sqrt{p^{0}_{l}p^{0}_{k}}}\,e^{-\ii(p_{l}x)+
    \ii(p_{k}y)
    +2\pi\ii {L}/\tilde{L}^{(kl)}}
    \left(n^{(l)}\otimes\na {k}{\phantom{j}}\right) U_{\alpha
k}U_{\alpha l}^{*}
\\[16pt] \times\displaystyle \left[\sqrt{\frac{p_{l}^{0}m_{l}}{p_{k}^{0}m_{k}}}
\left(1+\gamma^{0}\right)
+\sqrt{\frac{p_{l}^{0}+m_{l}}
{p_{k}^{0}+m_{k}}}\left(1-\gamma^{0}\right)\right]\!
(\gamma_{\mu}{p}^{\mu}_{k} - m_{k})(1 - \zeta\gamma ^5\gamma_{\mu} {S}_0^{\mu}(p_{k}) ).
\end{array}
\end{equation}

\section{Calculation of the pion decay probability}\label{A.2}

The expressions \eqref{26} after summation over the polarizations of the final particles can be easily reduced  to the following form
\begin{equation}\label{16.111}
\begin{array}{c}
\displaystyle
W_{\beta\alpha}^{L}=\frac{G^{2}_{\mathrm{F}}f_{\pi}^{2}}
{4(2\pi)^{6}k^{0}}\!\int\!\!
d^{4}x d^{4}y\!\int\!\! d^{4}qd^{4}p\,
\delta(p^{2}-M^{2}_{\beta})\delta(q^{2}-m^{2})
\\[4pt]\displaystyle\times
{\mathrm{Sp}}\left[(\gamma_{\mu}{p}^{\mu} - M_{\beta})\gamma_{\rho}(1+\gamma^5)(\gamma_{\nu}
{q}^{\nu} + m)\gamma_{\lambda}(1+\gamma^5)k^{\rho}k^{\lambda}\right]\\
\displaystyle \times\Bigg[\sum\limits_{k,l=1}^{3}e^{\ii(qx)[m_{k}/m
]-\ii(qy)[m_{l}/m ]+\ii((p-k)(x-y))+2\pi \ii
L/L^{(lk)}}\\[-8pt]\displaystyle \times
\left(m_{l}/{m}\right)^{3/2} \left(m_{k}/{m}\right)^{3/2} P_{\beta
l}P_{\beta k}^{*}U_{\alpha k}U_{\alpha l}^{*}\Bigg].
\end{array}
\end{equation}
\noindent Having calculated the  trace of $ \gamma $-matrices and
changed the integration variables
\begin{equation}\label{1300}
q^{\mu} \rightarrow q^{\mu} ({m}/{m_{kl}}), \quad  m_{kl}=({m_{k}+m_{l}})/2,
\end{equation}
\noindent we get
\begin{equation}\label{16.1111}
\begin{array}{c}
\displaystyle
W_{\beta\alpha}^{L}=\frac{2\,G^{2}_{\mathrm{F}}
f_{\pi}^{2}}{(2\pi)^{6}k^{0}}
\Bigg[\sum\limits_{k,l=1}^{3}
\frac{P_{\beta
l}P_{\beta k}^{*}U_{\alpha k}U_{\alpha l}^{*}}{\Delta^{3}_{lk}}\\ \displaystyle \times \!\!\int\!\! d^{4}\!q\,d^{4}\!p\,
\delta(p^{2}-M^{2}_{\beta})\delta(q^{2}-m^{2}_{kl})
\left(2(pk)(qk)-k^{2}(pq)\right)\\[1pt]
\displaystyle \times \!\!\int\!\!
d^{4}\!x \,d^{4}\!y\, e^{\ii(qx)[m_{k}/m_{kl}
]-\ii(qy)[m_{l}/m_{kl} ]+\ii((p-k)(x-y))+2\pi \ii
L/L^{(lk)}}\Bigg].
\end{array}
\end{equation}
\noindent  Here
\begin{equation}\label{130}
L^{(lk)}= \frac{4\pi |{\bf q}|}{m_{l}^{2}-m_{k}^{2}},\quad
 \Delta_{lk}=\frac{m_{k}+m_{l}}{2\sqrt{m_{l}m_{k}}}.
\end{equation}
\noindent We change the integration variables
\begin{equation}\label{x1}
\begin{array}{c}
\displaystyle (x^{0},{\bf {x}})\rightarrow \left(u^{0}x'^{0}+({\bf {u}}{\bf {x}'}),\,{\bf {x}'}+{\bf {u}}x'^{0}+\frac{{\bf {u}}({\bf {u}}{\bf {x}'})}{1+u^{0}}\right),\\[4pt] \displaystyle  (y^{0},{\bf {y}})\rightarrow \left(u^{0}y'^{0}+({\bf {u}}{\bf {y}'}),\,{\bf {y}'}+{\bf {u}}y'^{0}+\frac{{\bf {u}}({\bf {u}}{\bf {y}'})}{1+u^{0}}\right),
\end{array}
\end{equation}
\noindent where
\begin{equation}\label{x2}
u^{\mu}=q^{\mu}/m_{kl}.
\end{equation}
\noindent The Jacobian of transformation \eqref{x1} $ J = 1 $.
This change of variables yields the following expression in the
exponential
\begin{equation}\label{xx3}
\begin{array}{l}
\displaystyle \left(\frac{((k-p)q)}{m_{kl}}-m_{k}\right)x'^{0}-
\left(\frac{((k-p)q)}{m_{kl}}-m_{l}\right)y'^{0}\\[4pt]
-\displaystyle  \left(\left({\bf{k}}-{\bf{p}}-\frac{((k-p)q)+m_{kl}
(k^{0}-p^{0})}
{m_{kl}(m_{kl}+q^{0})}{\bf{q}}\right) \left({\bf {x}'}-{\bf {y}'}\right)\right).
\end{array}
\end{equation}
\noindent We change the variables
\begin{equation}\label{xx4}
\begin{array}{ll}
\displaystyle
{\bf {z}}'_{-}={\bf {x}'}-{\bf {y}'},& {\bf {z}}'_{+}=({\bf {x}}'+{\bf {y}}')/2,\\[4pt] z'^{0}_{-}=x'^{0}-y'^{0},& z'^{0}_{+}=(x'^{0}+y'^{0})/2.
\end{array}
\end{equation}
\begin{equation}\label{x4}
\;\;{\bf {z}}'_{-}=\left({\mathds{I}}-\frac{{\bf {q}}\otimes{\bf {q}}}{q^{0}(m_{kl}+q^{0})}\right){\bf {z}}''_{-}.
\end{equation}
\noindent The  Jacobian of transformation \eqref{x4} $ J'=
m_{kl}/q^{0} $. As a result, the expression in the
exponential takes the form
\begin{equation}\label{x3}
 \displaystyle \left(\frac{((k-p)q)}{m_{kl}}-m_{kl}\right)z'^{0}_{-} -
\left({m_{k}}-m_{l}\right)z'^{0}_{+} -  \left(\left({\bf{k}}-{\bf{p}}-\frac{k^{0}-p^{0}}
{q^{0}}{\bf{q}}\right) {\bf {z}}''_{-}\right).
\end{equation}

To get the probability of the process in a unit volume per unit
time we use  the standard Fermi ansatz.  We integrate
\eqref{16.1111} over $d^{4}z'_{+} $ in a finite four-dimensional
region $ V'T' $ and divide by the invariant volume $ V'T'= VT
$. As a result of this operation,  a factor
\begin{equation}\label{x40}
R=\frac{\sin \left(({m_{l}}-m_{k})T'/2\right)}
{\left({m_{l}}-m_{k}\right)T'/2}
\end{equation}
\noindent arises. It follows from Eq. \eqref{x1} that
\begin{equation}\label{x400}
T'= \frac{q^{0}T -({\bf{q}}{\bf{L}})}{m_{kl}}.
\end{equation}
\noindent Assuming ${\bf{L}}={\bf{v}}T, {\bf{v}}={\bf{q}}/q^{0}$, we get
\begin{equation}\label{x4000}
T'= \frac{m_{kl}}{q^{0}}T \equiv \frac{m_{kl}}{|{\bf {q}}|}L_{0},
\end{equation}
\noindent where $ L_{0} $ is the linear dimension of the region where the
pion decays. Hence
\begin{equation}\label{x40000}
R=\frac{\sin (\pi L_{0}/L^{(lk)})}{\pi L_{0}/L^{(lk)}}.
\end{equation}
The integration over $ d{\bf{z}}'_{-} $ yields the $\delta
$-functions associated with the momentum conservation
\begin{equation}\label{x5}
 (2\pi)^{3}\delta\left({\bf{k}}-{\bf{p}}-
 \frac{k^{0}-p^{0}}
{q^{0}}{\bf{q}}\right).
\end{equation}
\noindent It follows from Eq. \eqref{x5} that
\begin{equation}\label{x50000}
 {\bf{p}}={\bf{k}} -\frac{k^{0}-p^{0}}{q^{0}}{\bf{q}},
\end{equation}
\noindent and we have
\begin{equation}\label{x50}
\frac{((k-p)q)}{m_{kl}}-{m_{kl}}= \frac{m_{kl}}{q^{0}}(k^{0}-p^{0}-q^{0}).
\end{equation}
\noindent Therefore, using an integration variable
\begin{equation}\label{x500}
z^{0}_{-}=z'^{0}_{-}\frac{m_{kl}}{q^{0}}=z'^{0}_{-}J'
\end{equation}
\noindent and integrating over it, we obtain the  $ \delta
$-function associated with energy conservation
\begin{equation}\label{x5000}
2\pi\delta\left(k^{0}-p^{0}-q^{0}\right).
\end{equation}

Thus,
\begin{equation}\label{16.1112}
\begin{array}{c}
\displaystyle
W_{\beta\alpha}^{L}=\frac{2\,G^{2}_{\mathrm{F}}
f_{\pi}^{2}}{(2\pi)^{2}k^{0}}
\Bigg[\sum\limits_{k,l=1}^{3}\frac{P_{\beta
l}P_{\beta k}^{*}U_{\alpha k}U_{\alpha l}^{*}}{\Delta^{3}_{lk}}
\\ \times \displaystyle\int d^{4}qd^{4}p\,\frac{\sin (\pi L_{0}/L^{(lk)})}{\pi L_{0}/L^{(lk)}}e^{2\pi \ii
L/L^{(lk)}}\!\!\left(2(pk)(qk)-k^{2}(pq)\right)\\  \times\delta^{4}
\left(k-p-q\right)
\delta(p^{2}-M^{2}_{\beta})\,\delta(q^{2}-m^{2}_{kl})
\Bigg].
\end{array}
\end{equation}
\noindent The integral over the momentum variables coincides with
the one that appears  when calculating the probability of the
mass state creation.  Therefore,
\begin{equation}\label{16.1113}
\begin{array}{c}
\displaystyle
W_{\beta\alpha}^{L}=\frac{G^{2}_{\mathrm{F}}
f_{\pi}^{2}}{(2\pi)^{2}k^{0}}
\Bigg[\sum\limits_{k,l=1}^{3}\frac{P_{\beta
l}P_{\beta k}^{*}U_{\alpha k}U_{\alpha l}^{*}}{\Delta^{3}_{lk}}
\int d^{4}q\,\frac{\sin (\pi L_{0}/L^{(lk)})}{\pi L_{0}/L^{(lk)}}e^{2\pi \ii
L/L^{(lk)}}\, \\ [6pt] \times \displaystyle\left[M_{\beta}^{2}({m_{\pi}^{2}}-{M_{\beta}^{2}}+
m_{kl}^{2})+
m_{kl}^{2}({m_{\pi}^{2}}+{M_{\beta}^{2}}-m_{kl}^{2})
\right]\\ [2pt] \times \displaystyle\delta(q^{2}-m_{kl}^{2}) \delta(m_{\pi}^{2}-M^{2}_{\beta}+m_{kl}^{2}-2(kq))
\Bigg].
\end{array}
\end{equation}
\noindent  Assuming $ {\bf k} = 0 $ and neglecting $ m_{kl} $ in
comparison with the neutrino  energy we obtain the formula
\eqref{29}.

It should be noted that the change of variables  \eqref{28}
and limiting the  integration domain to the size of the area where
the reaction occurs is not a clearly defined operation, but rather
a prescription. However, such a procedure is not only justified by
physical considerations of Schwinger \cite{Schwinger}, but is also
mathematically consistent \cite{vladimirov}.

\end{document}